\documentclass[aps,prb,twocolumn,superscriptaddress,showpacs,floatfix]{revtex4}

\usepackage{amsmath}
\usepackage{amssymb}
\usepackage{epsfig}
\usepackage{multirow}
\usepackage{graphicx}
\newcommand {\dd}{{\mathrm d}}
\newcommand {\rr}{\mathbf{r}}
\newcommand {\drr}{{\mathrm d}\mathbf{r}}

\newcommand {\etal}{\begin{itshape}et al\end{itshape}.}

\newcommand{\suprm}[1]{\ensuremath{^{\mathrm{#1}}}}
\newcommand{\subrm}[1]{\ensuremath{_{\mathrm{#1}}}}

\begin{document}

\title{Solvent mediated interactions between model colloids and interfaces:\\
A microscopic approach}

\author{Paul\ Hopkins}
\email[]{Paul.Hopkins@bristol.ac.uk}
\affiliation{H.H.\ Wills Physics Laboratory, University of Bristol, Tyndall Avenue, Bristol BS8 1TL, UK}

\author{Andrew\ J. Archer}
\email[]{A.J.Archer@lboro.ac.uk}
\affiliation{Department of Mathematical Sciences, Loughborough University, Loughborough LE11 3TU, UK}

\author{Robert\ Evans}
\affiliation{H.H.\ Wills Physics Laboratory, University of Bristol, Tyndall Avenue, Bristol BS8 1TL, UK}

\date{\today}

\begin{abstract}
We determine the solvent mediated contribution to the effective potentials for model colloidal or nano- particles dispersed in a binary solvent that exhibits fluid-fluid phase separation. The interactions between the solvent particles are taken to be purely repulsive point Yukawa pair potentials. Using a simple density functional theory we calculate the density profiles of both solvent species in the presence of the `colloids', which are treated as external potentials, and determine the solvent mediated (SM) potentials. Specifically, we calculate SM potentials between (i) two colloids, (ii) a colloid and a planar fluid-fluid interface, and (iii) a colloid and a planar wall with an adsorbed wetting film. We consider three different types of colloidal particles: colloid A which prefers the bulk solvent phase rich in species 2, colloid C which prefers the solvent phase rich in species 1, and `neutral' colloid B which has no strong preference for either phase, i.e. the free energies to insert the colloid into either of the coexisting bulk phases are almost equal. When a colloid which has a preference for one of the two solvent phases is inserted into the disfavored phase at statepoints close to coexistence a thick adsorbed `wetting' film of the preferred phase may form around the colloids. The presence of the adsorbed film has a profound influence on the form of the SM potentials. In case (i) reducing the separation between the two colloids of type A leads to a bridging transition whereby the two adsorbed films connect abruptly and form a single fluid bridge. The SM potential is strongly attractive in the bridged configuration. A similar phenomenon occurs in case (iii) whereby the thick adsorbed film on colloid A and that at the planar wall, which prefers the same phase as colloid A, connect as the separation between the colloid and the wall is reduced. In both cases the bridging transition is accompanied, in this mean-field treatment, by a discontinuity of the SM force. On the other hand, for the same wall, and a colloid of type C, the SM potential is strongly repulsive at small separations. For case (ii), inserting a single colloidal particle near the planar fluid-fluid interface of the solvent, the density profiles of the solvent show that the interface distortion depends strongly on the nature of the colloid-solvent interactions. When the interface disconnects from the colloid there is, once again, a discontinuity in the SM force.
\end{abstract}


\maketitle

\section{Introduction}
\label{sec:part_inter_intro}

The effective forces between large bodies in a solvent, for example between suspended particles or between particles and container walls, are composed of contributions from the `direct' interactions, e.g. Coulomb and dispersion forces and also from `indirect' or solvent mediated interactions arising from the presence of the solvent. A significant part of colloid science consists of determining, and often tailoring, the effective interactions between colloidal particles.~\cite{barrat2003bcs,likos2001eis} In this paper we focus on a particular class of solvent induced interactions which arise from the adsorption of liquid films around large bodies. We consider large bodies suspended in a bulk solvent that exhibits coexisting fluid phases. The existence and thickness of the adsorbed `wetting' films depend strongly on the statepoint of the solvent. By changing the solvent temperature or concentration such that the fluid approaches coexistence, the thickness of the adsorbed film can increase and when two bodies with adsorbed films come sufficiently close these films can join together to form a fluid bridge. Such a mechanism generates strong attractions between the bodies that often lead to aggregation phenomena. Aggregation may be reversed by reversing the temperature or concentration change -- see Ref.~\cite{beysens1999wia} for an overview of the purported role of bridging in colloidal flocculation.

In the last decade or so interest has also grown in understanding the effective interactions between colloidal and nano- particles and interfaces, and between pairs of particles adsorbed at an interface.~\cite{bresme2007trn} Colloidal particles are often attracted to fluid interfaces and can stabilize emulsions in a similar way to surfactants.~\cite{aveyard1995lda,binks2002pss,aveyard2003ess} The adsorption of colloidal and nano-scale particles at interfaces is important for a number of industrial processes including foams, lubrication, adhesion, and stabilizing emulsions.~\cite{aveyard1995lda,binks2002pss,aveyard2003ess} The self assembly of nanoparticles and colloids adsorbed at fluid interfaces \cite{lin2003naa,lin2005naf} has been applied in many ways including creating solid `capsule' structures.~\cite{dinsmore2002csp} Particles at interfaces are also used in fundamental studies of the phase behaviour and physical properties of (quasi) two-dimensional fluids and crystals.~\cite{pieranski1980tdi,zahn2000dcm,bausch2003gbs, terao1999cqt} Colloidal nanoparticles may also be used as building blocks for materials with specific mechanical, optical and magnetic properties.~\cite{ozin2005nca}

Here we investigate the effective interactions between `colloidal' particles and fluid interfaces using a microscopic treatment of a model system where the fluid in which the `colloids' are dispersed, henceforth referred to as the `solvent', is composed of a binary mixture of soft-core particles which are much smaller than the `colloids'. These `solvent' particles interact via purely repulsive point Yukawa potentials which constitute a simple model for the interactions between charged particles -- see Refs.\ \cite{barrat2003bcs,hopkins2006pcf,hopkins2008iwp} and references therein. We showed previously~\cite{hopkins2006pcf} that for certain choices of mixing parameters this model solvent exhibits fluid-fluid phase separation and we focus our present study on state-points where the solvent is at, or close to, fluid-fluid coexistence. Our aim is to investigate the effective solvent mediated (SM) interaction between large bodies (`colloids') in the presence of the adsorbed films. In the following we consider three physical scenarios: (i) the effective interaction between two colloids of the same type immersed in the bulk solvent; (ii) that between a single colloid and the planar interface between two coexisting fluid phases; and (iii) that between a single colloid and a planar wall which is covered by a thick adsorbed film of solvent. In each case we determine the equilibrium density profiles of the binary solvent in the presence of the colloids and from these profiles we calculate the SM interactions. We use a simple microscopic density functional theory (DFT), shown to be reliable for bulk fluids, to calculate the density profiles and grand potential of the solvent in the external potential of the fixed colloids. We describe the interaction between the colloids and the solvent particles by a hard-core potential plus a purely repulsive Yukawa tail that models crudely the effective interaction between a hard body carrying a surface charge and solvent particles with the same sign. Although our system may be considered to be a toy model for charged colloidal particles, we believe that our results should provide insight into the behavior of any system of colloidal or nano- particles that are dispersed in a solvent exhibiting fluid-fluid phase separation.

In the previous study~\cite{hopkins2006pcf} we showed that the pair correlation functions of the bulk fluid at states close to coexistence are reasonably well described by a simple random phase approximation (RPA). Subsequently, using a density functional theory that generates the RPA, we showed that the model fluid wets completely a hard-wall and exhibits a pre-wetting transition slightly away from coexistence.~\cite{hopkins2008iwp} Furthermore, by adding a repulsive tail to the wall potential we found that the location and extent of the pre-wetting line in the phase diagram can be changed. Although the model solvent is very simple, its bulk and interfacial phase behavior mimics that in more sophisticated models and has the important advantage that it can be studied using the computationally inexpensive RPA based DFT.

We consider three different colloids that vary in their affinity for one of the coexisting bulk fluid phases. Colloid type A has a very strong preference for the phase rich in species 2, colloid B is almost `neutral' but has a very weak preference for the phase rich in species 2, and colloid C has a strong preference for the phase rich in species 1. When colloids A and C are inserted into their unfavored phase they adsorb a thick layer of their preferred phase. We find that the propensity of an isolated colloid immersed in the bulk phase to adsorb thick films dominates the behavior of the solvent when the colloids are brought together, or brought to interfaces, and therefore determines to a large extent the nature of the SM interactions between bodies. In particular we find that for two colloids in bulk the presence of thick adsorbed films leads to a long-ranged and strongly attractive SM interaction: as the two colloids are brought together the adsorbed films join together abruptly to form a bridge between the colloids thereby minimizing the interfacial contribution to the grand potential. If in isolation the colloids do not adsorb thick films then the SM potential is much shorter ranged but is still attractive due to depletion effects. For a colloid that strongly prefers to be in one of the phases the effective interaction with the fluid-fluid interface is highly asymmetric with the global minimum of the grand potential, that determines the colloid's equilibrium position, occurring in bulk, far away from the interface. For the `neutral' colloid the global minimum of the SM interaction is in the center of the interface where the colloid intersects the interface. For the third scenario, a single colloid near a planar wall, we find that if the colloid prefers the phase adsorbed at the wall then there is a long-ranged, attractive SM interaction between the wall and the colloid. If the colloid does not strongly prefer either phase it seeks a location that intersects the interface between the thick adsorbed film and the bulk fluid and the interaction is attractive but not as strong, nor as long-ranged, as the previous case. On the other hand, if the colloid prefers the bulk phase then the SM potential will be repulsive, and relatively short-ranged. 

The use of DFT for studying the wetting behavior of a solvent in the presence of an isolated big spherical particle that exerts a (spherical) external potential on the solvent atoms is, of course, well established see e.g.~\cite{bieker1998wcs,stewart2005wdc,evans2004ncc} and references therein. However, the behavior of the solvent in the presence of two fixed particles or a particle and a wall is not as widely studied. These more complex geometries generally require considerably more computational effort. In the following we mention several different approaches that have been used to investigate these problems using DFT. One of earliest studies of the SM interaction between two spherical particles exerting dispersion forces made use of an interface potential approach where the solvent density profile between liquid and gas phases is modeled by a sharp-kink approximation.~\cite{dobbs1992ccb,dobbs1992ccp} By solving for the position of the interface around a pair of colloids at decreasing separations an abrupt crossover from unconnected to connected liquid films, termed capillary condensation by the authors of~\cite{dobbs1992ccb,dobbs1992ccp}, was found. In Ref.~\cite{bauer2000wie} a more sophisticated free-energy functional was employed but the sharp-kink interface approximation was retained. The authors investigated the morphological transition in some detail and also compared the SM interaction potential to the bare interaction arising from dispersion forces between the pair of spheres.

An alternative DFT approach developed by Roth \etal~\cite{roth2000dph} makes use of the potential distribution theorem which provides a formally exact expression for the SM potential in terms of the one-body direct correlation function between the inserted particle and the solvent. This in turn depends only on the inhomogeneous solvent density profile around a single, isolated (spherical) particle {\it before} the second particle is inserted. Given a suitable density functional, the direct correlation function and the spherically symmetric density profile of the solvent around a single colloid are easily determined and it is straight forward to calculate the SM potential between two large particles. The advantage of this approach is that one must only calculate a solvent density profile that has spherical symmetry. The disadvantage is that one requires a DFT reliable for highly asymmetric mixtures. This so called insertion method has been used to calculate the SM potential for various types of hard and soft-core model particles.~\cite{roth2000dph,archer2003smi,archer2002mts}

In Ref.~\cite{archer2005smi} Archer \etal~compared the results of the insertion method to those of a `brute-force' DFT approach for calculating the SM potential between two large Gaussian soft-core particles immersed in a bulk binary solvent of smaller soft-core particles. Note that we refer to the procedure that calculates the solvent density profiles around two fixed colloids treated as external potentials as the `brute-force' method, in contrast to the insertion method \cite{roth2000dph} that requires only a calculation of the solvent density profiles around a single fixed colloid. The authors found that for certain solvent statepoints close to fluid-fluid coexistence the large particles could adsorb a thick film of the wetting `phase'. Using the brute force approach, it was shown that on reducing the separation between the two large particles the adsorbed films would connect abruptly forming a fluid bridge. Such a bridging transition leads, at mean-field level, to a discontinuity in the SM force between two large particles and in the bridged phase the potential is long-ranged and strongly attractive. Results from the insertion method, using the same RPA based DFT but now for a ternary mixture, shown to be accurate for solvent statepoints away from the binodal, could not account for the bridging transition. However, the insertion method does predict, albeit qualitatively, the occurrence of long-ranged SM potentials when the thick adsorbed films are present.~\cite{archer2005smi}

A number of other studies have used brute-force DFT methods to calculate the SM potential.  Stark \etal~\cite{stark2004cci} used Landau-de Gennes theory to investigate capillary bridging in the case of two large hard spherical colloids immersed in a bulk isotropic liquid crystal host at statepoints close to the isotropic-nematic phase boundary. Grodon~\cite{grodonthesis} studied a model colloid-polymer mixture adsorbed between two large colloidal spheres. Andrienko \etal~\cite{andrienko2004cba} used Landau theory to calculate density profiles and force distance curves for a solvent adsorbed between a large sphere and a planar wall. They found a first-order capillary bridging transition and investigated its dependence on the radius of the sphere. Cheung and Allen~\cite{cheung2006slc,cheung2007lcm,cheung2008fbc} have used an Onsager (second virial) DFT for hard rods to investigate the effective interactions between two cylindrical colloids, and between a cylindrical colloid and a wall, in a liquid-crystal host. At suitable statepoints in the bulk isotropic phase, regions of nematic order can form at the wall and around the cylinders. The adsorbed nematic regions can then form a bridge, in an analogous fashion to the above examples, giving rise to strongly attractive interactions.

We are not aware of studies that use DFT to study the interaction between a large particle and a fluid interface although we are aware of studies that use self-consistent field-theory (SCFT) to investigate the interactions between particles and interfaces in self-organized block co-polymer structures.~\cite{thompson2001pmc,kim2009pjn}

There is a growing body of computer simulation investigations concerned with the adsorption of individual spherical, and non-spherical, particles at fluid interfaces.~\cite{bresme1999csw,bresme1999nll,faraudo2003spa,bresme2007ota,cheung2009ion,bresme2007trn} In general the size of the nanoparticles is of the same order as the fluid particles and thick adsorbed layers are not present or are not considered. Much is made of line tension contributions to the free energy of immersion.~\cite{bresme1999csw,cheung2009ion}

Our paper proceeds as follows: In Sec.~\ref{sec:pi_review} we recall some of the basic background to the thermodynamics of solvation, colloid-colloid and colloid-interface interactions. In Sec.~\ref{sec:theory} we outline the model system and in Sec.\ \ref{sec:densfunc} we describe our density functional theory approach. Sec.~\ref{sec:results} describes the results of this investigation. Finally in Sec.~\ref{sec:discuss} we discuss the results and make some concluding remarks. Note that we frequently use the term `colloid' for the particle or particles inserted into the solvent. However, the inserted particle has an effective radius of $4.5\lambda^{-1}$ whereas that of the solvent particles is about $0.5\lambda^{-1}$. Such length scales are much more appropriate to those associated with inserting nanoparticles into solvents.

\section{Background: Some key concepts}
\label{sec:pi_review}

\subsection{Thermodynamics of solvation}
\label{sec:col_fi_pre}
Before describing our microscopic DFT approach for calculating SM potentials, we first recall some of the thermodynamics relevant to a colloid (spherical particle) immersed in a fluid. We consider a grand canonical system enclosed within a volume $V$ containing several different species of solvent particles, that is coupled to a reservoir which fixes the temperature $T$ and the set of chemical potentials $\{\mu_i\}$ for the different solvent species $i$ in the system. Later in this paper we consider the specific case of a solvent composed of two different species of particles (i.e.\ $i=1$ or 2), but for now we leave the number of components undefined. At equilibrium, the grand potential of the uniform solvent is
\begin{equation}
\Omega(V,T,\{\mu_i\})=-PV,
\label{eq:Omega_empty}
\end{equation}
where $P$ is the pressure and we have assumed that there are no interfacial contributions. If we insert a spherical particle (colloid) with radius $R$, the grand potential is
\begin{equation}
\Omega(V,T,\{\mu_i\},R)=-P(V-\frac{4}{3}\pi R^3)+4\pi R^2\gamma_{p 
\alpha}(R),
\label{eq:Omega_particle}
\end{equation}
where $(V-\frac{4}{3}\pi R^3)$ is now the volume that is accessible to the solvent (which we denote phase $\alpha$) and $\gamma_{p\alpha}(R)$ is the surface tension (surface excess grand potential per unit area) between the colloidal particle $p$ and phase $\alpha$.~\cite{footnote1} Note that when $R$ is large, and in the absence of wetting films:
\begin{equation}
\gamma_{p\alpha}(R)=\gamma_{p \alpha}(\infty) \left(1-\frac{2\delta}{R} 
+ \cdots \right),
\end{equation}
where $\delta$ is the Tolman length, i.e. a microscopic length scale of order the radius of the solvent particles. \cite{rowlinson2002mtc,evans2004ncc,stewart2005wdc} If we subtract Eq.\ \eqref{eq:Omega_empty} from Eq.\ \eqref{eq:Omega_particle}, we obtain the following expression for the excess grand potential for inserting the particle into the system (phase $\alpha$):
\begin{equation}
\Omega\suprm{ex}_\alpha=\frac{4}{3}\pi R^3P+4\pi R^2\gamma_{p\alpha}(R).
\label{eq:Omega_ex}
\end{equation}

We now consider the case when the chemical potentials $\{\mu_i\}$ and the temperature $T$ are such that we observe two phase coexistence between phase $\alpha$ and a second phase which we denote $\beta$. In this case, the grand potential of the system is
\begin{equation}
\Omega(V,T,\{\mu_i\})=-PV+\gamma_{\alpha \beta}A,
\label{eq:Omega_empty_2phase}
\end{equation}
where $\gamma_{\alpha \beta}$ is the surface tension for the (planar) interface between phases $\alpha$ and $\beta$ and $A$ is the area of this  
interface. If we insert the colloid into phase $\alpha$, far away from the interface, then the excess grand potential is still given by Eq.\ \eqref{eq:Omega_ex}, but if we insert the colloid into phase $\beta$, again at a point far away from the interface, then the excess grand potential is now
\begin{equation}
\Omega\suprm{ex}_\beta=\frac{4}{3}\pi R^3P+4\pi R^2\gamma_{p\beta}(R),
\label{eq:Omega_ex_beta}
\end{equation}
where $\gamma_{p\beta}(R)$ is the surface tension between the colloidal particle $p$ and phase $\beta$.

When the two surface tensions $\gamma_{p \alpha}$ and $\gamma_{p \beta}$ are equal, then the colloid does not favor one phase over the other. However, this is a special case and more generally the surface of the colloid will favor one phase over the other. Suppose it favors phase $\alpha$, i.e. $\gamma_{p \alpha}<\gamma_{p \beta}$. There is an upper bound on the value $\gamma_{p\beta}$ can take, since in the case when the surface of the colloid strongly favors the $\alpha$ phase, on inserting the colloid into the coexisting $\beta$ phase, a film of the favored $\alpha$ phase forms around the colloid. Thus, the upper bound on $\gamma_{p \beta}$ is given by $\gamma_{p \beta} \leq \gamma_{p \alpha}+\gamma_{\alpha \beta}$. The equality occurs in the limit that the radius $R \to  \infty$, and when the surface is completely wet by phase $\alpha$. For a colloid of radius $R$ having this kind of surface, we find that the excess grand potential for inserting it into the $\beta$ phase is:
\begin{equation}
\Omega\suprm{ex}_\beta\approx\Omega^{ex}_\alpha+4\pi (R+l)^2\gamma_{\alpha \beta} 
(R+l),
\label{eq:Omega_ex_beta_wet}
\end{equation}
where $\Omega\suprm{ex}_\alpha$ is given by Eq.\ \eqref{eq:Omega_ex}, $l$ is the thickness of the film of phase $\alpha$ covering the surface of the colloid and $\gamma_{\alpha \beta}(R)$ is the surface tension of the spherical fluid-fluid interface between the wetting $\alpha$ phase and the bulk $\beta$ phase. Note that in Eq.~\eqref{eq:Omega_ex_beta_wet} we have omitted the interaction between the two interfaces $p\alpha$ and $\alpha \beta$. This result also holds when the solvent $\beta$ phase is not exactly at coexistence: changing the chemical potentials so that the mixture is (not too far) off coexistence simply reduces the wetting film thickness $l$. However, there is an additional contribution proportional to $l$ arising from the fact that the $\alpha$ phase is metastable. \cite{stewart2005wdc,evans2004ncc}

\subsection{Solvent mediated potentials in a bulk fluid}
\label{sec:smp_rev}

We now consider inserting two colloids into the bulk fluid. When the pair of colloids are far apart (separated by a distance $h \to \infty$), then the insertion free energy $\Omega\suprm{ex}(h \to \infty)$ is simply twice the result in either Eq.\ \eqref{eq:Omega_ex} or Eq.\ \eqref{eq:Omega_ex_beta}, depending into which phase $\alpha$ or $\beta$ the colloids are inserted. However, in the case when we insert the two colloids separated by a finite distance, the insertion free energy $\Omega\suprm{ex}(h)$ now depends on the distance between the centres of the colloids $h$ and also on the bulk phase into which the colloids are inserted. The SM potential is defined as follows:
\begin{equation}
W(h)=\Omega\suprm{ex}(h)-\Omega^{ex}(h\to\infty),
\label{eq:smp_00}
\end{equation}
i.e.\ it is the difference between the grand potential when the colloids are far from one another ($h \to \infty$) and when they are separated a distance $h$. The total effective potential between the two colloids is $U(h)+W(h)$ where $U(h)$ is the `bare' or direct interaction.

When the colloids are inserted into the preferred $\alpha$ phase $W(h)$ is fairly short ranged. Due to depletion forces \cite{barrat2003bcs} $W(h)$ is typically attractive when the colloids are close to contact. Generally $W(h)$ has a range determined by the bulk correlation length in the solvent. However, if we insert the two colloids into phase $\beta$ close to or at bulk coexistence, where both colloids are surrounded by a thick adsorbed wetting film of phase $\alpha$, then $W(h)$ can become much greater in magnitude, much longer ranged and strongly dependent on the conformations of the thick films surrounding the colloids. When the colloids are far apart the adsorbed films do not interact, i.e.\ when $h$ is large $W(h)\simeq 0$. As the colloids are brought closer together, the surrounding wetting films begin to interact and if the colloids are sufficiently close the system can minimize the interfacial area between the wetting $\alpha$ and bulk $\beta$ phases (and therefore minimize the grand potential) by bridging the gap between the colloids and creating a single interface. Without explicitly solving for the minimal interfacial shape as a function of $h$ (see Eq.\ \eqref{eq:S_approx} below for an approximate analytic expression for $S(h)$), we may write the excess grand potential for inserting the two colloids into the $\beta$ phase as follows [c.f.\ Eq.\ \eqref{eq:Omega_ex_beta_wet}]:
\begin{equation}
\Omega\suprm{ex}_\beta(h) \simeq 2\Omega\suprm{ex}_\alpha+8\pi(R+l)^2\gamma_{\alpha \beta}(R+l) S(h),
\label{eq:Omega_ex_beta_wet_2}
\end{equation}
where $S(h)$ is the ratio of total surface area of the adsorbed wetting film(s) when the particles are a distance $h$ apart to that when they are infinitely far apart. In Eq.\ \eqref{eq:Omega_ex_beta_wet_2} we have assumed that the average surface tension (excess grand potential per unit area) of the $\alpha$--$\beta$ interface is equal to $\gamma_{\alpha \beta}(R+l)$ for all surface conformations. When the adsorbed films do not interact, $S(h)=1$. However, when the colloids become close enough for the films to connect and form a bridge, then $0.5<S(h)\leq 1$. Bridge formation leads to a discontinuity in the first derivative of $S(h)$ at $h=h_{br}$. For $h<h_{br}$, $S(h)$ decreases smoothly as the colloids are brought to contact. Using Eq.\ \eqref{eq:Omega_ex_beta_wet_2}, we can approximate the SM potential as follows:
\begin{eqnarray}
W(h)&\simeq&8\pi(R+l)^2\gamma_{\alpha \beta}(R+l) (S(h)-1) \notag \\
&\simeq&-8\pi R^2\gamma_{\alpha \beta}[1-S(h)],
\label{eq:W_approx}
\end{eqnarray}
where we have assumed in the second line that $\gamma_{\alpha \beta}(R+l)\simeq\gamma_{\alpha \beta}(\infty)\equiv\gamma_{\alpha \beta}$, the surface tension of the planar $\alpha$--$\beta$ interface, and we have used the fact that $R\gg l$. This means that it is the quantity ${\cal A}=R^2\gamma_{\alpha \beta}$ which determines the value of $W(h)$ when the colloids are surrounded by thick wetting films and close to contact, $h\simeq h_{c}$. Since typically  the interfacial tension $\gamma_{\alpha\beta} \sim k_BT/\sigma^2$, where $k_B$ is Boltzmann's constant and $\sigma$ is a molecular length scale (the size of the solvent atoms/molecules), we find that ${\cal A} \sim k_BT(R/\sigma)^2$. Thus for even modest sized colloids of radius 10nm, since $\sigma \simeq$0.2nm for a typical solvent, the gain in free energy in bringing the colloids close to contact can be large; in this case, $W(h\subrm{c}) \sim -10^3$ to $-10^4k_BT$. It is for this reason that one should expect colloids to aggregate strongly when the solvent is near to phase coexistence in cases where the surface of the colloids is such that they are wet by the coexisting phase.

\subsection{Interaction between a colloid and an interface}
\label{sec:nondeform_inter}

For colloids dispersed in a solvent which is at bulk phase coexistence (we continue to denote the two coexisting phases $\alpha$ and $\beta$) one can consider the effective interaction between the colloid and the interface between the two phases. One generally finds that large colloidal particles are strongly attracted to the interface and become strongly bound to it. The reason for this is that if one places the colloid within the interface, so that the interface intersects the colloid, then the area of the interface between fluid phases $\alpha$ and $\beta$ is reduced by an amount $\simeq \pi R^2$ and so there is a change in the free energy
\begin{equation}
\Delta \Omega \simeq -\pi R^2 \gamma_{\alpha\beta}
\label{eq:omega_interface}
\end{equation}
on moving the colloid from the bulk fluid into the interface. Again we find it is the quantity ${\cal A}=R^2\gamma_{\alpha \beta}$ which determines the magnitude of $\Delta \Omega$. For colloids of radius 10nm, the change in grand potential $\Delta \Omega \sim -10^3$ to $-10^4k_BT$.

Going beyond this very rough approximation, we can consider a single large spherical colloid embedded within a non-deformable planar fluid-fluid interface: the excess grand potential as a function of the variable $\bar{z}_0=z_0/R$, where $z_0$ is the distance of the centre of the colloid from the plane of the interface, is given by \cite{aveyard1996pwa, bresme2007trn}:
\begin{eqnarray}
\Omega\suprm{ex}=\Omega_\alpha\suprm{ex}
+2\pi R^2\bigg[(1-\bar{z}_0)(\gamma_{p\beta}-\gamma_{p\alpha})\notag \\
-\frac{1}{2}(1-\bar{z}_0^2)\gamma_{\alpha \beta}+\frac{\tau}{R} \sqrt{1-\bar{z}_0^2}\bigg],
\label{eq:Fint_0}
\end{eqnarray}
when the colloid is within the interface, $-1\leq \bar{z}_0\leq 1$. $\Omega_\alpha\suprm{ex}$ is given by Eq.\ \eqref{eq:Omega_ex} and $\tau$ is the line tension. The line tension was originally introduced by Gibbs to describe the excess free energy associated with the line where three phases meet.~\cite{rowlinson2002mtc} This quantity may be positive or negative and experimental measurements of the line tension have proved to be difficult. \cite{amirfazli2004stp} For large $R$ the contribution from the line tension to Eq.\ \eqref{eq:Fint_0} will be small compared to the contributions from the surface tensions. However, for nanoparticles line tension effects can be significant; see Ref.~\cite{bresme2007trn}, and references therein.

By considering the interface to be a deformable membrane, one can go beyond Eq.\ \eqref{eq:Fint_0} and include the contributions to $\Omega\suprm{ex}$ from the bending and stretching of the interface. \cite{bresme2007trn, oettel2005eci} This approach has also been extended to determine the effective potential between two colloids that are within an interface.~\cite{oettel2008cif} However, this interesting problem is beyond the scope of the present study, though we return briefly to it in Sec.~\ref{sec:discuss}.

\section{Model System}
\label{sec:theory}

\subsection{Solvent}

\begin{figure}[tp]
\centering
\includegraphics[width=8.5cm]{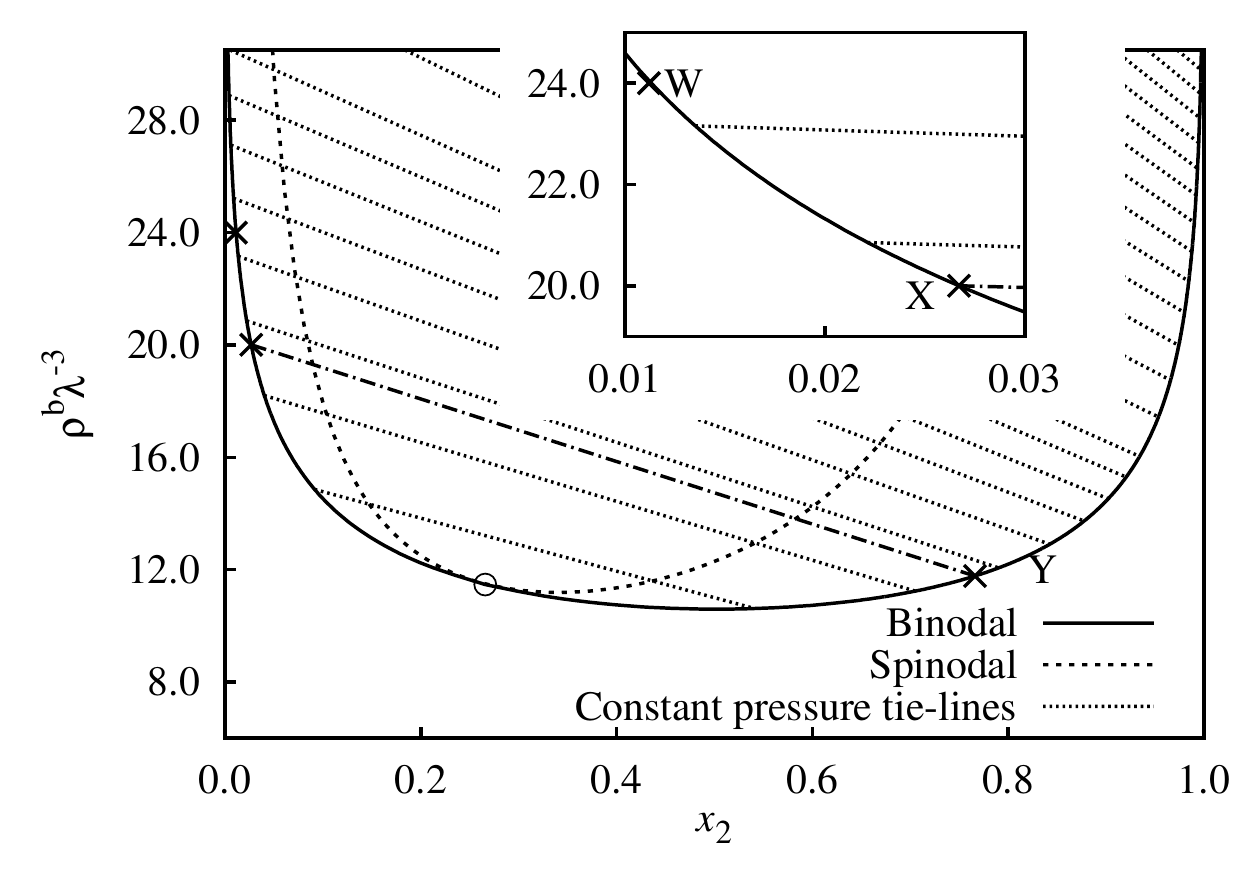}
\caption{\label{fig:phased}
Bulk phase diagram of the binary point Yukawa fluid (our solvent) with parameters $M_{11}=1$, $M_{12}=2.2$, $M_{22}=4$ and temperature $k_BT/\epsilon=1$. $\rho^b$ is the total bulk density and $x_2$ is the concentration of species 2. The de-mixed region is bounded by the binodal (solid line) and meets the spinodal (dashed line) at the critical point, indicated by $\circ$. The straight (dotted) tie-lines connect coexisting state-points with pressures $\beta \lambda^{-3} P=150$ to $500$, in increments of $50$, and then from $500$ to $1900$ in increments of $100$ (from bottom to top). The crosses mark three particular state-points: the co-existing pair X and Y at the pressure $\beta \lambda^{-3} P=233$, which are joined by the dash-dot tie-line and the point W that lies very close to coexistence at $\rho^b\lambda^{-3}=24.0$. Further details concerning these three state points, at which we display various results below, are given in Table \ref{tab:state_points}. The inset displays a magnification of the region around points W and X.}
\end{figure}

The solvent is a binary mixture of particles that interact via purely repulsive, point Yukawa pair potentials
\begin{equation}
 \phi_{ij}(r)=\frac{\epsilon M_{ij}}{4\pi}\frac{\exp(-\lambda r)}{\lambda r},
\label{eq:yuk_pot_def}
\end{equation}
where the parameter $\epsilon>0$ sets the energy scale, $M_{ij}$ is the interaction magnitude between species $i$ and $j$, and $\lambda$ is an inverse length scale. In Ref.~\cite{hopkins2006pcf} we showed that when $M_{12}>\sqrt{M_{11}M_{22}}$ and the bulk fluid total number density $\rho^b\equiv \rho_1+\rho_2$ is sufficiently high, then the system phase separates into two fluid phases, one phase rich in particles of species 1 and the other phase rich in particles of species 2. The phase diagram for the mixture with parameters $M_{11}=1$, $M_{22}=4$, $M_{12}=2.2$ and temperature $k_BT/\epsilon=1$, calculated using the random phase approximation (RPA) for the Helmholtz free energy (see Ref.~\cite{hopkins2006pcf} and Sec.\ \ref{sec:densfunc} below for further details) is displayed in Fig.~\ref{fig:phased}. Clearly this system exhibits fluid-fluid phase separation. Indeed this is the main reason for its selection as model solvent in this study. Another important reason for studying this model fluid is that in Ref.~\cite{hopkins2008iwp} we showed that when this system is at coexistence there can be complete wetting of the planar hard-wall by the phase rich in species 2 accompanied out of coexistence by a first-order surface (pre-wetting) phase transition from a thin to a thick adsorbed wetting film. We showed that this was also the case for a planar hard wall potential augmented with repulsive exponential or Yukawa potentials. We found that the location and extent of the pre-wetting line depends strongly on the parameters of the wall potential.~\cite{hopkins2008iwp} These properties allow us to use this microscopic model to investigate the influence of wetting films on solvent mediated potentials between colloidal particles.

\subsection{Colloids}
\label{subsec:part_int}

We model the interaction of the colloids with the solvent particles $i=$1,2 via a hard-core plus repulsive Yukawa-tail pair potential:
\begin{equation}
 \phi^{\mathrm{sc}}_i(r)=
\begin{cases}
  \infty & 0<r < R \\
\epsilon b_i\frac{\exp(-\lambda (r-R))}{\lambda (r-R)} & R < r,
\end{cases}
\label{eq:testpart}
\end{equation}
where $R$ is the radius of the colloids and the parameter $b_i>0$ determines the magnitude of the repulsive tail interaction with solvent particles of species $i$. The value of the ratio $b_1/b_2$ is important for determining whether the surface of the colloid favors solvent particles of species 1 or particles of species 2. In this study we choose both $b_1$ and $b_2$ to be fairly large in order to `soften' the boundary of the hard-core. If the potentials in Eq.\ \eqref{eq:testpart} change rapidly over a short distance (which is the case when both $b_1$ and $b_2$ are small), and the potentials are defined on a Cartesian grid this can lead to numerical errors in the solvent density profiles, which are defined on the same Cartesian grid. All the results presented in this paper are for the case when the radius of the colloids $R=4\lambda^{-1}$. The effective radius of the solvent particles is roughly between $0.2\lambda^{-1}$ -- $0.5 \lambda^{-1}$ for the densities of interest here. We estimate this radius from the size of the `correlation hole' in the radial distribution functions $g_{ij}(r)$ -- see Ref.\ \cite{hopkins2006pcf} for examples of $g_{ij}(r)$ for this system. A radius $R=4\lambda^{-1}$ is sufficiently large for the colloid to be covered by a thick adsorbed `wetting' film when the ratio $b_1/b_2$ is such that the colloid favours one solvent species over the other and the solvent is at a state point near to coexistence. In this study, we set $b_1=10$ and $b_2=10$, 20 and 30, thus defining three different types of colloids, A, B and C. We find that colloids of type A, with $b_2=10$, have a strong preference for solvent particles of the `bigger' species 2. Colloids of type B, with $b_2=20$, have a very weak preference for species 2; B colloids are essentially `neutral' colloids. Colloids of type C, with $b_2=30$, have a strong preference for species 1, the `smaller' solvent particles. Note that we do not define the `bare' interaction potential $U(h)$ between the colloids. This quantity does not enter into our calculation of the solvent mediated potential $W(h)$.

\subsection{Wall-Solvent Interfaces}
\label{subsec:ext_pot}

We model the potentials exerted by a planar wall on the solvent particles as follows:
\begin{equation}
 V^{\mathrm {sys}}_i(z)=
\begin{cases}
 \infty & z<0 \\
a_i\epsilon\exp(-\lambda z)/(\lambda z) & z\geq 0,
\end{cases}
\label{eq:yukwall}
\end{equation}
where the parameter $a_i>0$ determines the magnitude of the repulsive tail interaction with solvent particles of species $i$. In a manner entirely analogous to the way the colloid potentials in Eq.\ \eqref{eq:testpart} influence the solvent, the value of ratio $a_1/a_2$ is important for determining whether the wall favors solvent particles of species 1 or solvent particles of species 2, i.e.\ the precise values of the parameters $a_1$ and $a_2$ determine whether there is a wetting film of the coexisting phase adsorbed at the wall when the bulk phase is brought to coexistence and also determine the location and extent in the phase diagram of the pre-wetting phase transition line.~\cite{hopkins2008iwp} In this study we set $a_1=a_2=1$. For these values we find that there is a weak preference for particles of species 2 at the wall, so that when the bulk phase rich in particles of species 1 is brought near to coexistence, we find that there is a thick wetting film of the phase rich in species 2 adsorbed at the wall. Any wetting transition occurs at densities above those shown in Fig.~\ref{fig:phased}. These values of $a_1$ and $a_2$ are sufficiently large that both solvent fluid density profiles vary sufficiently slowly for there not to be any numerical errors due to the discretisation of the density profiles.

\section{Density Functional Theory and its Implementation}
\label{sec:densfunc}

Here we present a brief description of the DFT approach that we use to calculate the solvent density profiles around the colloid(s) and to calculate thermodynamic quantities such as $\Omega\suprm{ex}$. For a more complete account of DFT see e.g.\ Refs.~\cite{evans1979nlv,evans1992fif,hansen2006tsl}.  For a fluid composed of $\nu$ different species of particles one can construct a functional $\Omega[\{\rho_i\}]$ of the set of one-body density profiles $\{\rho_i(\rr)\}$, $i=1\dots \nu$. The minimum value of this functional is equal to the thermodynamic grand potential of the system, i.e.\ $\Omega=\min(\Omega[\{\rho_i\}])$, and the set of density profiles which minimize the functional $\Omega[\{\rho_i\}]$ are the equilibrium fluid density profiles satisfying the following set of Euler-Lagrange equations:
\begin{equation}
\frac{\delta \Omega[\{\rho_i\}]}{\delta\rho_i(\rr)}=0,
\label{eq:EL_0}
\end{equation}
which may be solved to obtain the set of equilibrium fluid density profiles. The grand potential functional may be written as:
\begin{equation}
 \Omega[\{\rho_i\}]=F[\{\rho_i\}]-\sum_{i=1}^{\nu}\int\drr\rho_i(\rr)(\mu_i-V_i(\rr)),
\label{eq:omega}
\end{equation}
where $F[\{\rho_i\}]$ is the intrinsic Helmholtz free energy functional and $V_i(\rr)$ is the external potential acting on particles of species $i$. The intrinsic Helmholtz free energy functional may be separated into a sum of two contributions: $F[\{\rho_i\}]=F_{\mathrm{id}}[\{\rho_i\}]+F_{\mathrm{ex}}[\{\rho_i\}]$ where the first term is the Helmholtz free energy of an ideal-gas
\begin{equation}
F_{\mathrm{id}}[\{\rho_i\}] = \sum_{i=1}^{\nu} k_BT \int\drr\rho_i(\rr)(\ln(\Lambda_i^3\rho_i(\rr))-1),
\end{equation}
where $\Lambda_i$ is the thermal de Broglie wavelength of particles of species $i$, and the second term is the excess contribution due to the particle interactions.

The solvent that we consider is a binary mixture ($\nu=2$) of soft-core particles interacting via the pair potentials in Eq.\ \eqref{eq:yuk_pot_def}. For this system, at the state points of interest, it is sufficient to employ a simple mean-field approximation for the excess part of the Helmholtz free energy \cite{hopkins2008iwp}:
\begin{equation}
F_{\mathrm{ex}}[\{\rho_i\}] = \frac{1}{2}\sum_{i,j=1}^2\int \drr \int \drr'\rho_i(\rr)\rho_j(\rr')\phi_{ij}(|\rr-\rr'|).
\label{eq:rpa_func}
\end{equation}
This functional generates the RPA for the pair direct correlation functions: $c_{ij}(|\rr-\rr'|)=-\beta\frac{\delta^2 F_\mathrm{ex}[\{\rho_i\}]}{\delta\rho_i(\rr)\delta\rho_j(\rr')}=-\beta\phi_{ij}(|\rr-\rr'|)$, where $\beta=(k_BT)^{-1}$. \cite{likos2001eis, archer2001bgc}

In order to simplify the computation of the density profiles it is convenient to reduce the range of the solvent pair potentials $\phi_{ij}(r)$ by cutting and shifting these, i.e.\ we replace the solvent pair potentials $\phi_{ij}(r)$ by $\phi\suprm{cas}_{ij}(r)$, which are defined as follows:
\begin{eqnarray}
 \phi_{ij}\suprm {cas} (r) & = &
\begin{cases}
\theta \left[\phi_{ij}(r)-\phi_{ij}(r_c)\right] & r\leq r_c \\
0 & r > r_c,
\end{cases}
\label{eq:yuk_cas}
\end{eqnarray}
where $\theta \simeq1$ is a dimensionless factor. If we set $\theta=1$, then the system with the cut and shifted (CAS) potential does not have the same bulk phase diagram as the fluid with the full range potential $\phi_{ij}(r)$ (one is shifted slightly with respect to the other). However, if we choose the value of $\theta$ so that
\begin{equation}
\int_0^\infty\dd r r^2\phi_{ij}(r)=\int_0^{r_c}\dd r r^2\phi\suprm{cas}_{ij}(r),
\label{eq:cas_a_def}
\end{equation}
then within the RPA, the bulk phase diagram for the fluid with the CAS potential is identical to the phase diagram of the system with the full range potential. This makes comparisons with existing results more straightforward. In the present study we set the cut-off distance $r_c=4\lambda^{-1}$. This is large enough so as to not alter significantly the correlation functions of the fluid.

The colloids are treated as fixed external potentials, which must be added to any other external potentials, to give
\begin{equation}
V_i(\rr)=V_i^{\mathrm {sys}}(\rr)+\sum_{k}\phi^{\mathrm{sc}}_i(|\rr-\rr_k|)
\label{eq:vext_sum}
\end{equation} 
where $V_i^{\mathrm {sys}}(\rr)$ is given by Eq.\ \eqref{eq:yukwall} (when no wall is present, $V_i^{\mathrm {sys}}(\rr)=0$), $\phi^{\mathrm{sc}}_i(r)$ is given by Eq.\ \eqref{eq:testpart} and $\rr_k$ is the location of the centre of colloid $k$. In the present study we only consider situations that have cylindrical symmetry, i.e.\ one colloid plus a planar interface or an isolated pair of colloids, and so it is natural to characterize points in space by the coordinates $(z,r)$, where $z$ is the axial distance along the $z$-axis, which passes through the center of the colloid(s) and is perpendicular to the plane of the interfaces (when interfaces are present), and $r$ is the radial distance from the $z$-axis. We denote the location of the center of colloid $k$ along the $z$-axis by $z_k$.

The solvent density profiles are calculated numerically by solving Eqs.\ \eqref{eq:EL_0} self consistently using a simple Picard iteration algorithm. We use fast Fourier transforms to evaluate the convolution integrals in the axial direction, but the radial convolutions are integrated directly. This allows us to use a grid with a smoothly varying distance between grid-points in the radial direction. Close to the colloids we use a small grid spacing $\simeq0.1\lambda^{-1}$ in order to accurately determine the rapidly varying density profiles at these points, but further out, where the profiles vary slowly, we use a larger grid spacing $\simeq0.3\lambda^{-1}$ in order to increase the computation box size without any loss of performance.

In calculating the solvent density profiles around the colloid in the fluid-fluid interface it is necessary to fix the density profiles on the boundary of the system far away from the colloid along the $r$-axis to be equal to those in the unperturbed state, when the colloid is absent, with the interface fixed at a certain value of $z$:
\begin{equation}
\rho_i(z,r=r_L)=\rho_i(z,r\to\infty),
\label{eq:R_bc}
\end{equation} 
where $r_L$ is the size of the computation region. If this is not done, the entire interface simply translates to the value of $z$ where the free energy is the global minimum rather than giving the constrained free energy minimum where the colloid is constrained to be a certain distance from the plane of the interface. For the particle in the fluid interface calculations we set $r_L\simeq70\lambda^{-1}$. For the case of a colloid at a wall, provided that the size of the computation region is sufficiently large, the density profiles satisfy Eq.~\eqref{eq:R_bc} automatically, since the location of the interface between the wetting film and the bulk fluid is determined by the properties of the wall and the chemical potentials.

In order to calculate the surface tension (excess grand potential) $\gamma$ associated with the individual interfaces, and compare with results such as Eq.~\eqref{eq:W_approx}, we use the equilibrium fluid density profiles together with the following expression:
\begin{equation}
\gamma= \frac{\Omega+PV}{A}=\frac{1}{A}\int\dd \rr \left( \omega[\{\rho_i(\rr)\}]+P\right)
\label{eq:gamma_def}
\end{equation}
where $A$ is the area of the interface, $\omega[\{\rho_i(\rr)\}]$ is the grand free energy density [see Eq.\ \eqref{eq:omega}], and $P=-\omega[\{\rho_i(\infty)\}]$ is the bulk fluid pressure.

\section{Results of Calculations}
\label{sec:results}

\subsection{Interfaces in the absence of colloids}
\label{subsec:cas_bbpd}

Before examining the solvent mediated interactions between the colloid and the interfaces, we first describe briefly the properties of the solvent at the interfaces {\em without} any colloids present. Full details of the solvent interfacial behavior can be found in Ref.\ \cite{hopkins2008iwp}; here we merely recall the properties relevant to the present study.

\begin{figure}[t]
\centering
\includegraphics[width=8.5cm]{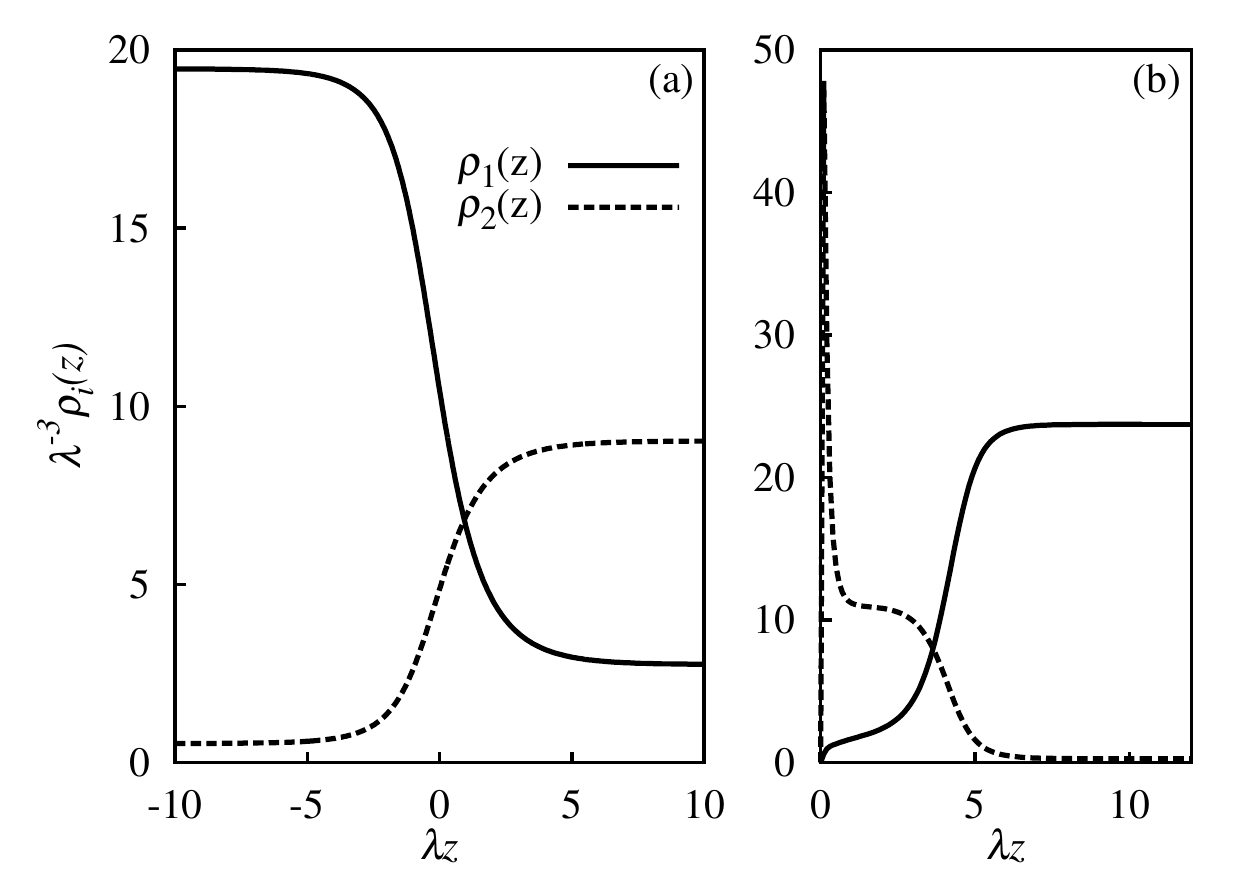}
\caption{\label{fig:wall_profiles}
Solvent density profiles $\rho_i(z)$ for two inhomogeneous systems: a) The planar fluid-fluid interface between coexisting phases at state-points X (rich in species 1) and Y (rich in species 2) in Fig.\ \ref{fig:phased}. b) The solvent density profiles at a planar wall with potentials given by Eq.\ \eqref{eq:yukwall}, for the bulk fluid at state point W in Fig.\ \ref{fig:phased}. These profiles show that there is a thick wetting film rich in solvent particles of species 2 adsorbed at the wall. Note that the density profiles of both species vanish at the wall.}
\end{figure}

In Fig.\ \ref{fig:wall_profiles}(a) we display the solvent density profiles at the planar fluid-fluid interface between the coexisting state points $X$ and $Y$, which have a bulk fluid pressure $\beta P \lambda^{-3}=233$. The values of the total density and concentration at each of these two state-points are listed in Table \ref{tab:state_points}. The density profiles are monotonic functions of $z$ and the (10\%-90\%) width of the interface is $4.2\lambda^{-1}$. These density profiles are typical of a binary system that is at a state point not too far removed from the bulk fluid critical point.

\begin{table}[t]
\begin{center}
\begin{tabular*}{0.9\columnwidth}{@{\extracolsep{\fill}}ccccc}
System & Labels & $\rho^b\lambda^{-3}$ & $x_2$ & $\beta\gamma\lambda^{-2}$ \\
\hline\hline
Thick film at wall  & W & 24.0 & 0.0112 & -116 \\
\hline
\multirow{2}{*}{Fluid-fluid interface} & X & 20.0 & 0.0267 &\multirow{2}{*}{5.46} \\
& Y & 11.8 & 0.766 & \\
\end{tabular*}
\caption{\label{tab:state_points} Two systems are considered: (i) state point W in Fig.\ \ref{fig:phased}, where the concentration $x_2$ is slightly less than the value at coexistence $x\subrm{2,coex}=0.01132$ and the planar wall is covered by a thick adsorbed wetting film rich in particles of species 2, and (ii) the fluid-fluid interface between co-existing phases at state points X and Y in Fig.\ \ref{fig:phased}. $\gamma$ is the excess grand potential per unit area (surface tension) for these interfaces and is calculated using Eq.\ \eqref{eq:gamma_def}.}
\end{center}
\end{table}

For the bulk fluid at state point W (see Fig.\ \ref{fig:phased}) in contact with a planar wall interacting with the solvent via the potentials in Eq.\ \eqref{eq:yukwall}, with $a_1=a_2=1$, we find that the wall is covered by a thick adsorbed wetting film rich in particles of species 2 -- see the density profiles displayed in Fig.~\ref{fig:wall_profiles}(b). The thickness of the wetting film is determined by the separation of the bulk state point from the binodal; the closer the state point is to the binodal, the thicker the wetting film. \cite{hopkins2008iwp} State point W is chosen because here the wall is covered by a relatively thick film with a well-defined interface boundary; recall that as one approaches the critical point, the width of the interface between the adsorbed film and the bulk fluid grows and ultimately diverges, as does the thickness of the adsorbed film. In Fig.~\ref{fig:wall_profiles}b we also see that there is a marked peak in the density profile for species 2, corresponding to a strongly adsorbed layer of particles at the wall. Beyond this layer, the thick wetting film extends out a distance $\simeq 5\lambda^{-1}$ from the wall.

\subsection{A single colloid in the bulk solvent}
\label{subsec:single_colloid}

\begin{figure}[t]
\centering
\includegraphics[width=8.5cm]{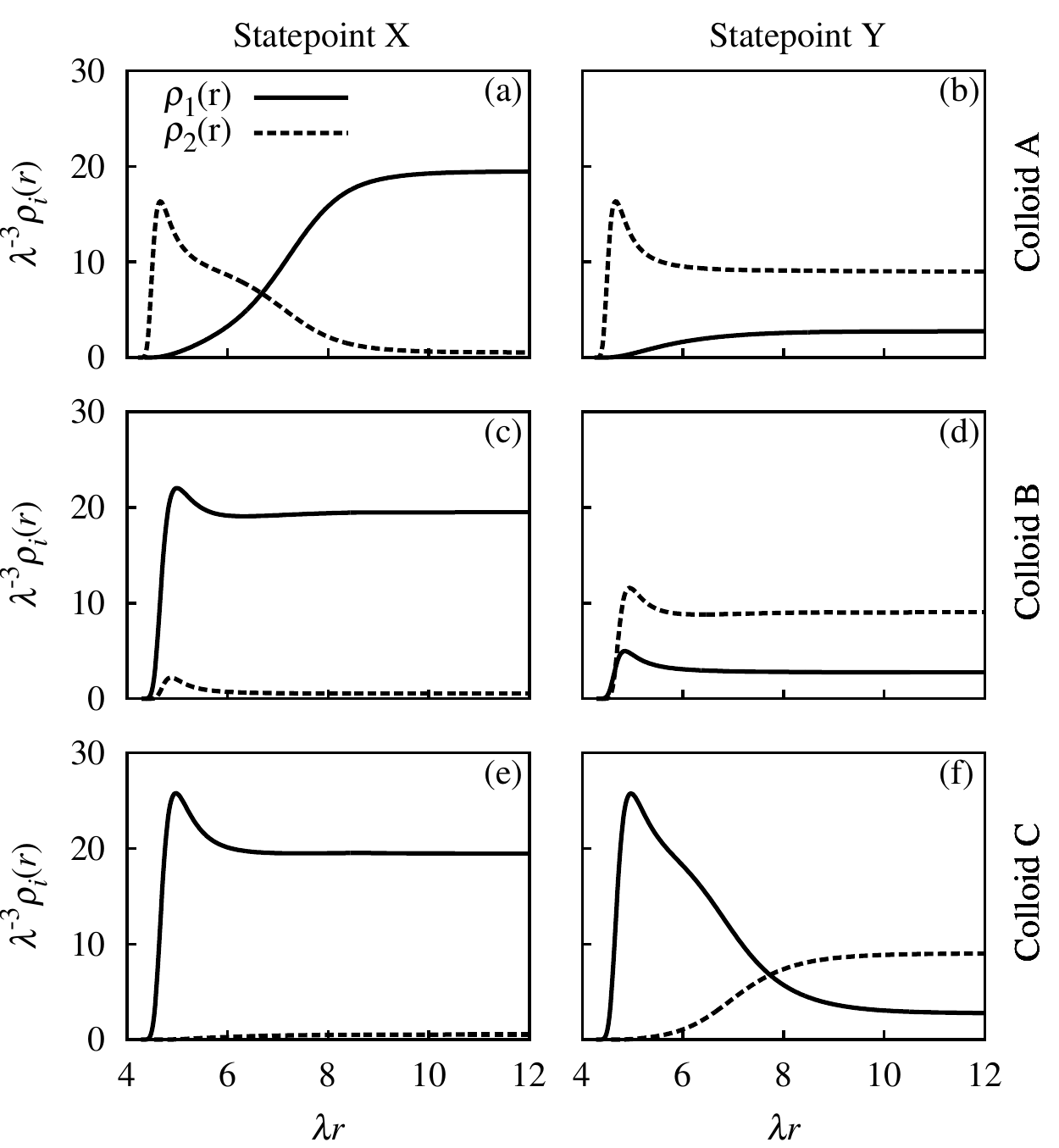}
\caption{\label{fig:colloid_profiles}
The solvent density profiles $\rho_i(r)$ around colloids A, B and C of radius $R=4\lambda^{-1}$ at the state-points X and Y in Fig.\ \ref{fig:phased} (see also Table \ref{tab:state_points}). $r$ is the distance from the centre of the colloid. See the text for a discussion of these density profiles.}
\end{figure}

We now describe the properties of the solvent around a single colloid immersed in a bulk phase that is at coexistence. Using the solvent-colloid pair potentials and the parameters defined in Sec.~\ref{sec:theory}B, we calculate the density profiles of the solvent around a single colloid at the coexisting state-points X and Y. Our results are displayed in Fig.\ \ref{fig:colloid_profiles}. At both state points, we see in Fig.\ \ref{fig:colloid_profiles}(a) and (b) that the surface of colloid A strongly prefers solvent particles of species 2; the density of species 1 particles at the surface of the colloid is much lower than that of species 2. This leads to the adsorption of a thick film of the coexisting phase rich in species 2 when colloid A is inserted in the bulk fluid at state-point X. Recall that the thickness $l(R)$ of this wetting film is determined by the curvature (i.e.\ radius $R$) of the surface of the colloid. \cite{dietrich12pta} In the limit that the colloid radius $R\to \infty$, the thickness of the wetting film of the coexisting phase diverges (becomes macroscopically large): for short-ranged forces of the type considered here $l\sim\ln(R)$. \cite{evans2004ncc} Colloid B has a weak preference for particles of species 2, but in Fig.\ \ref{fig:colloid_profiles}(c) and (d) we see only a small increase in the density of both species near the surface of colloid B; there are no thick wetting films. We see in Fig.\ \ref{fig:colloid_profiles}(e) and (f) that the behavior of the solvent around colloid C is essentially the opposite of the behavior around colloid A: particles of species 1 are strongly adsorbed at the surface, so that when colloid C is immersed at state-point Y, there is a thick adsorbed wetting film of the coexisting phase rich in species 1 particles around the colloid. From the density profiles we estimate an effective colloid radius, $R^*$, by taking the distance where the density of the strongly adsorbed species is $10\%$ of its maximum value. For all colloids, in both phases, this occurs at approximately $r=R^*=4.5\lambda^{-1}$. We also compute the grand potential $\Omega$ for colloids A, B and C within the bulk solvent phases at points X and Y. This allows us to quantify the preference of the colloids to be in either bulk phase by calculating the difference in grand potentials, $\Delta\Omega=\Omega\suprm{Y}-\Omega\suprm{X}$. Colloid A has a very large negative value, $\Delta \Omega=-4870k_BT$, indicating that it strongly prefers to be in bulk phase Y. Colloid B has small negative value, $\Delta \Omega=-403k_BT$, indicating a slight preference for phase Y. The grand potential difference for colloid C is $\Delta \Omega=4610k_BT$. This large positive value indicates that colloid C had a strong preference for phase X. Note that by phase X (Y) we mean the bulk phase at state-point X (Y). 

\subsection{A pair of colloids in the bulk solvent}
\label{sub:twocolls}

\begin{figure}[t]
\centering
\includegraphics[width=8.5cm]{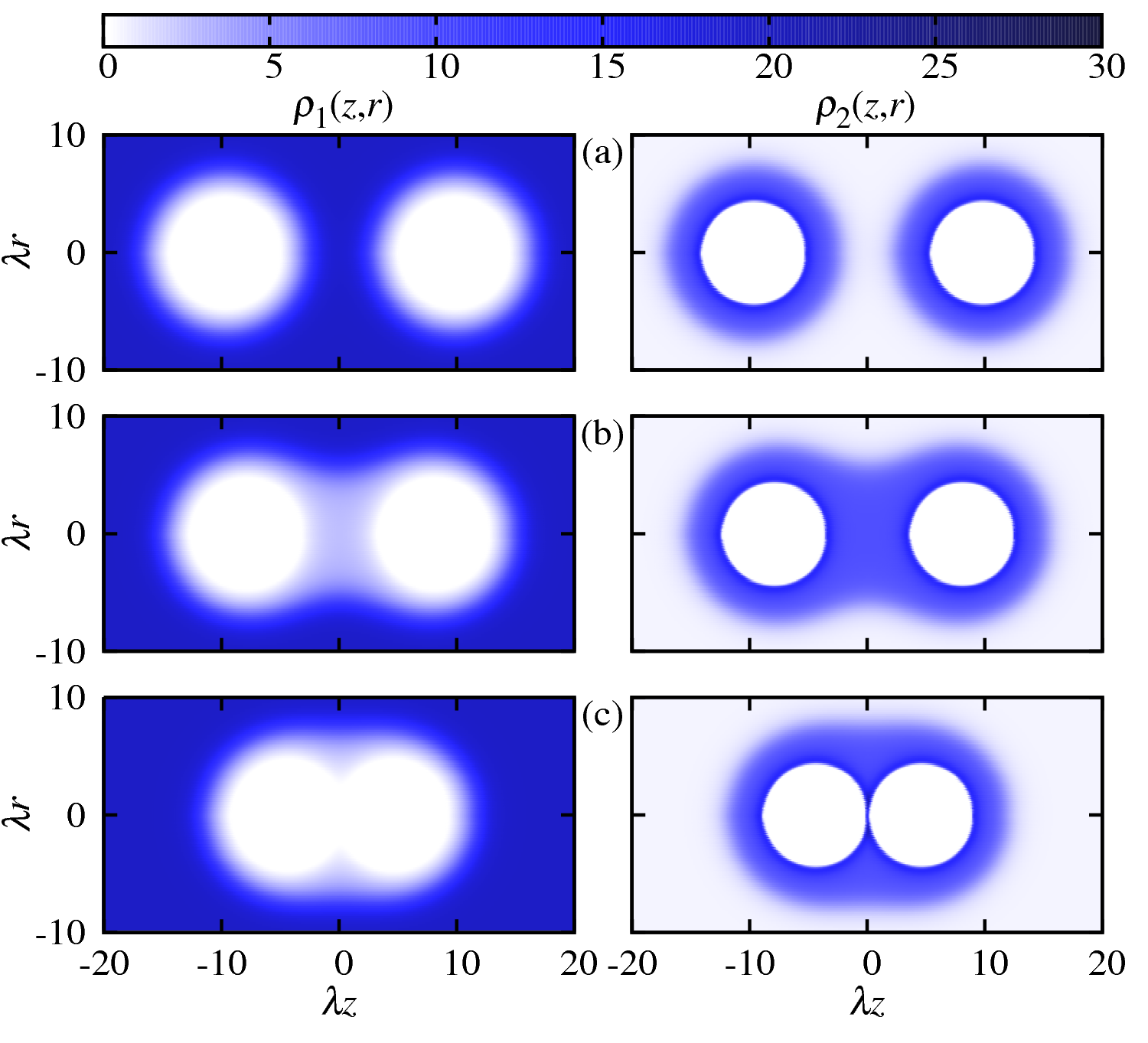}
\caption{\label{fig:two_part_dens_profsA}
(Colour online) The solvent density profiles $\lambda^{-3}\rho_i(z,r)$ at the bulk state point X, around a pair of type A colloids of radius $R=4\lambda^{-1}$, separated a distance (a) $h=19.5\lambda^{-1}$, (b) $h =16\lambda^{-1}$ and (c) $h=9.0\lambda^{-1}$. The left hand plots are the density profiles for species 1 and the right hand plots for species 2. In (a) we see the two colloids are covered by thick wetting films rich in species 2, but they are far enough apart that the films do not connect. In (b) the colloids are sufficiently close that the films connect to form a bridge between the two colloids. In (c) the colloids are close to contact. The solvent mediated potentials corresponding to these profiles are denoted by $\Box$ in Fig.~\ref{fig:two_part_omega_bar}.}
\end{figure}

\begin{figure}[t]
\centering
\includegraphics[width=8.5cm]{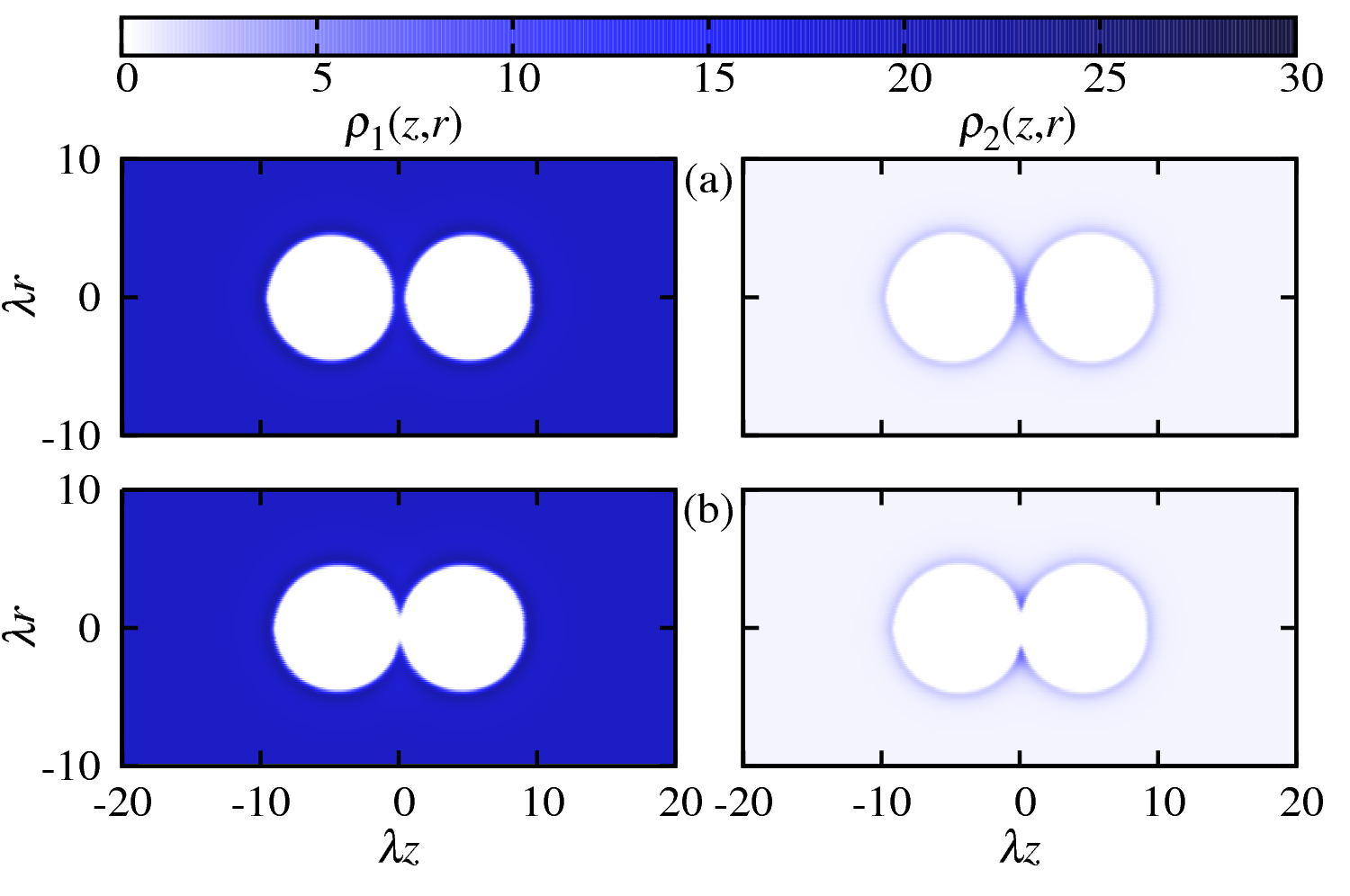}
\caption{\label{fig:two_part_dens_profsB}
(Colour online) The solvent density profiles $\lambda^{-3}\rho_i(z,r)$ at the bulk state point X, around a pair of type B colloids of radius $R=4\lambda^{-3}$, separated a distance (a) $h=10\lambda^{-1}$ and (b) $h=9\lambda^{-1}$. The left hand figures are the density profiles for species 1 and the right hand figures for species 2. There is an adsorbed layer of both solvent species around the colloids but there are no thick wetting films. The solvent mediated potentials corresponding to these profiles are denoted by $\triangle$ in Fig.~\ref{fig:two_part_omega_bar}.}
\end{figure}

 \begin{figure}[t]
\centering
\includegraphics[width=8.5cm]{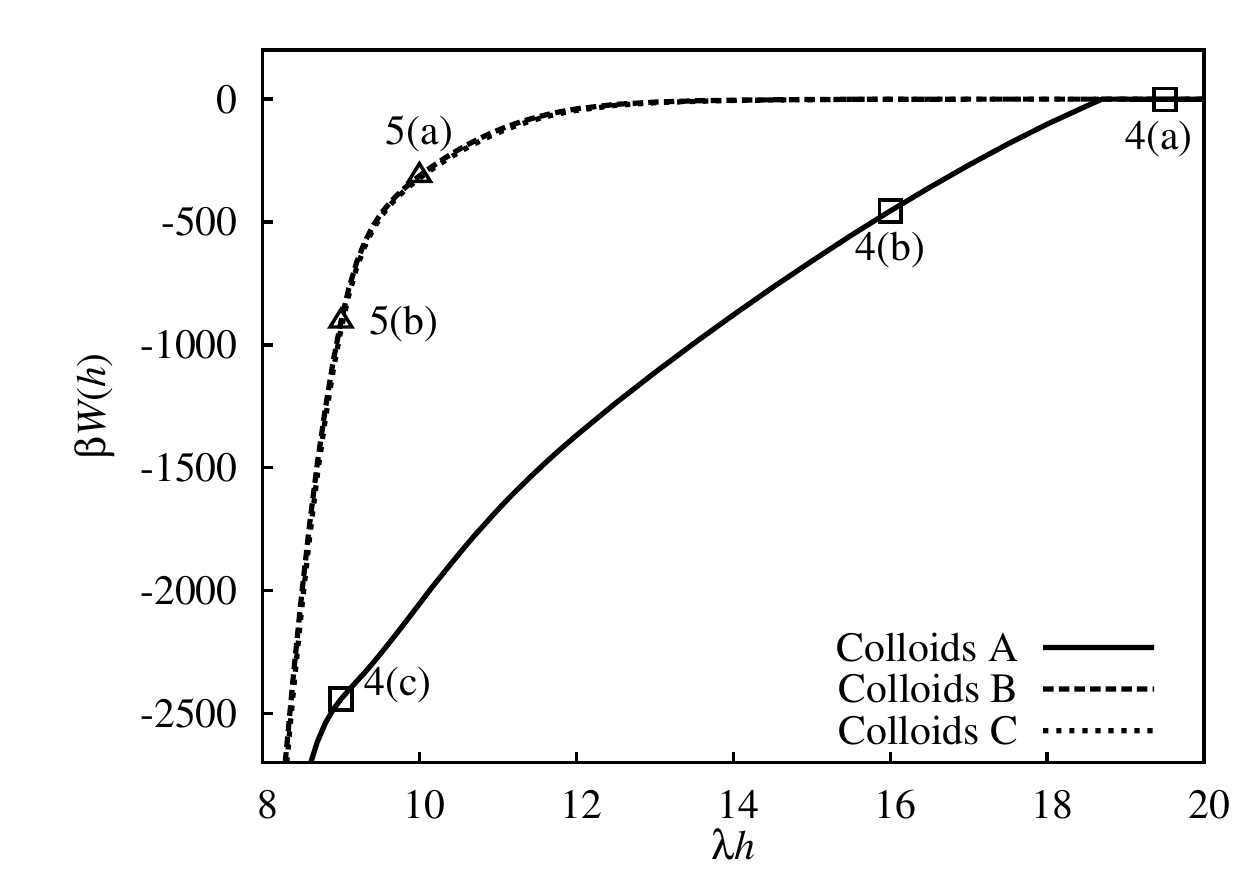}
\caption{\label{fig:two_part_omega_bar}
The solvent mediated potential $W(h)$, where $h=z_0-z_1$ is the distance between the centers of the colloids, for pairs of colloids of the same type, A, B, and C. The solvent is at state-point X in Fig.\ \ref{fig:phased}. For pairs of type B and C colloids, $W(h)$ is a smooth continuous function of $h$ and is almost identical for both types of colloids. It is also strongly attractive: $W(h=9\lambda^{-1})=-905k_BT$. For type A colloids, $W(h)$ is even more strongly attractive, $W(h=9\lambda^{-1})=-2442k_BT$, and much longer ranged. We also observe a `kink' (discontinuity in the first derivative) in $W(h)$ at $h\subrm{br}\simeq18.7\lambda^{-1}$, where the thick films around the A colloids first connect to form a bridge. The symbols $\Box$ and $\triangle$ carry labels referring to Figs.~\ref{fig:two_part_dens_profsA} and \ref{fig:two_part_dens_profsB} which display the solvent density profiles for the particular separation.} 
\end{figure}

We now describe the behavior of the solvent for two colloids separated by a distance $h$ in the bulk fluid at statepoint X. The SM potential between them, $W(h)$, is defined in Eq.\ \eqref{eq:smp_00}. In Fig.\ \ref{fig:two_part_dens_profsA} we display the solvent density profiles around two colloids of type A at various separations $h$ between the centers of the colloids and in Fig.\ \ref{fig:two_part_omega_bar} we display the SM potential $W(h)$ obtained from these density profiles. As discussed in the previous subsection, we find that the colloids are covered by a thick adsorbed wetting film that is rich in particles of species 2. For large separations $h\gtrsim 19 \lambda^{-1}$ the films do not interact with each other, see Fig.~\ref{fig:two_part_dens_profsA} and the potential $W(h)=0$. As the separation $h$ is decreased, the wetting films around the two colloids begin to influence one another and when the colloids are brought to a separation $h=h\subrm{br}=18.7\lambda^{-1}$ the thick films surrounding the pair of colloids join to form a bridge between the colloids composed of the wetting phase rich in particles of species 2; see Fig.~\ref{fig:two_part_dens_profsA}(b). For all separations $h< h_{br}$, the bridge remains between the two colloids. In Fig.~\ref{fig:two_part_omega_bar} we see that at $h=h_{br}$, the separation at which the adsorbed films switch from the unbridged to the bridged state, the first derivative of $W(h)$ changes discontinuously, i.e.\ there is a jump in the solvent mediated force $f=-{\rm d} W/{\rm d}h$.~\cite{footnote2} For $h<h_{br}$, $W(h)$ decreases rapidly as $h$ is decreased. We say more below about $W(h)$ when there is bridging.

In Fig.\ \ref{fig:two_part_dens_profsB} we display the solvent density profiles around a pair of type B colloids immersed in the solvent phase X for the separations $h=9\lambda^{-1}$ and $h=10\lambda^{-1}$. These `neutral' colloids are not covered by any thick wetting films. As the two B colloids are brought together, the solvent density profiles change continuously with $h$ and we do not find any bridging. Comparing the density profiles in Fig.\ \ref{fig:two_part_dens_profsB}(a) for $h=10\lambda^{-1}$ with those in Fig.\ \ref{fig:two_part_dens_profsB}(b) for $h=9\lambda^{-1}$, we see that as the pair of colloids are brought close together the solvent particles are expelled from the region directly between the two colloids. The SM potential $W(h)$ corresponding to this case is displayed in Fig.\ \ref{fig:two_part_omega_bar}. We find that $W(h)$ is strongly attractive (near to contact, $W(h=9\lambda^{-1})=-905k_BT$) and is fairly short ranged. When the two colloids are brought close to one another the expulsion of the solvent particles from the region between them results in fewer solvent particles being directly at the surface of the colloids. This in turn means that the energy penalty for solvent particles being close to the colloids, resulting from the repulsive tails of the potentials $\phi_i^{sc}(r)$, is reduced. This lowering of the potential energy then leads to a lowering of the free energy of the system when the colloids are closer together and so $W(h)$ is attractive. Below $h\simeq9\lambda^{-1}$ $W(h)$ decreases extremely rapidly as the strongly repulsive parts of the colloid-solvent potentials start to overlap.

In Fig.\ \ref{fig:two_part_omega_bar} we also display the solvent mediated potential between a pair of type C colloids. Type C colloids, like type B colloids, are not covered by any thick wetting films. We find that the solvent mediated potential $W(h)$ between a pair of type C colloids is almost indistinguishable from the same quantity for a pair of type B colloids for this state point X. The reason that the colloid B and colloid C solvent mediated potentials are so similar is that the two types of colloids  differ only in the value chosen for the parameter $b_2$ in Eq.\ \eqref{eq:testpart}, the magnitude of the interaction of the colloids with species 2 particles. At state point X, the density of the species 2 particles is small and so the potential energy contribution to the free energy from the interactions of the colloids with the species 2 solvent particles is also small.

If we compare in Fig.\ \ref{fig:two_part_omega_bar} the solvent mediated potential $W(h)$ for the case when there are thick wetting films around the colloids (type A) and those when there are no thick films (type B or C colloids), it is tempting to argue that by subtracting one potential from the other, one would be left with the contribution due solely to the thick wetting films present in the former case. We believe this argument to be cogent, at least at a qualitative level, on the basis of the following: Subtracting the value of the solvent mediated potential $W(h_c)$ near contact at $h=h_c=9\lambda^{-1}$ for type C colloids from the same quantity for type A colloids gives us an estimate for the contribution to the solvent mediated potential due to the fluid bridge, i.e. $W\subrm{br}(h=h_c)\simeq -2440k_BT+900k_BT= -1540k_BT$. We now return to Eq.\ \eqref{eq:W_approx} which is an expression obtained from macroscopic thermodynamic arguments for the contribution to $W(h)$ arising from the presence of thick wetting films. In Ref.\ \cite{archer2005smi} the following approximate expression for the ratio $S(h)$ of surface areas was obtained by assuming that the surface of the fluid bridge may be approximated by the surface generated by rotating the arc of a circle around the $z$-axis:
\begin{eqnarray}
S(h)&=&\frac{w h( L^2-(h/2-w)^2)^{1/2}}{4L(h/2-w)^2}\arcsin\left( \frac{h/2-w}{L} \right) \notag \\
&&+\frac{(L+h/2-w)}{2L} - \frac{w^2}{2L(h/2-w)},
\label{eq:S_approx}
\end{eqnarray}
where $L=R+l$ and $2w$ is the width along the $z$-axis of the bridge section -- see Ref.\ \cite{archer2005smi} for further details. $w$ is treated as a variational parameter, i.e.\ one selects the value of $w$ that minimizes $S(h)$. Using this expression for $S(h)$, together with the value $L=R+l=7.5 \lambda^{-1}$ (a rough estimate for the wetting film thickness based on inspecting the density profiles in Fig.\ \ref{fig:colloid_profiles}(a)) and the value for the surface tension of the liquid-liquid interface $\gamma$ listed in Table \ref{tab:state_points}, we obtain from Eq.\ \eqref{eq:W_approx} the value $W(h_c)\simeq -1620k_BT$ which is close to the value $W_{br}(h_c)\simeq -1540k_BT$ estimated above. Eqs.\ \eqref{eq:W_approx} and \eqref{eq:S_approx} together also predict that bridging occurs when the colloids are at a distance $h\simeq17.2\lambda^{-1}$, in fairly close agreement with the full DFT result of $h_{br}=18.7\lambda^{-1}$. These observations lead us to conclude that the approximation for $W(h)$ in Eq.\ \eqref{eq:W_approx}, taken together with a good approximation for $S(h)$, such as that in Eq.\ \eqref{eq:S_approx}, provide a fairly reliable description of the solvent mediated potential. Recall that Eq.\ \eqref{eq:W_approx} was derived from macroscopic arguments, so it is perhaps surprising that the present microscopic DFT results for a system where the colloids are only one order of magnitude bigger in size than the solvent particles are well accounted for by this equation.

The results presented in this subsection are for the colloids immersed in the solvent at state-point X. However, similar results are obtained at state-point W which is slightly off bulk coexistence -- see Table~\ref{tab:state_points}. Furthermore, the results obtained here for colloids of type A in the solvent at state point X are qualitatively the same as those for colloids of type C at state point Y, since in both cases the colloids are covered in thick films of the coexisting phase. Moreover the behavior of the systems with colloids of type A in the solvent at state point Y and colloids of type C in the solvent at state point X are also similar. The solvent around two colloids of type B behaves in much the same way in both bulk phases X and Y leading to similar solvent mediated potentials.

\subsection{Interaction between a single colloid and the fluid-fluid interface}
\label{subsec:ffi}

 \begin{figure}[t]
 \centering
 \includegraphics[width=8.5cm]{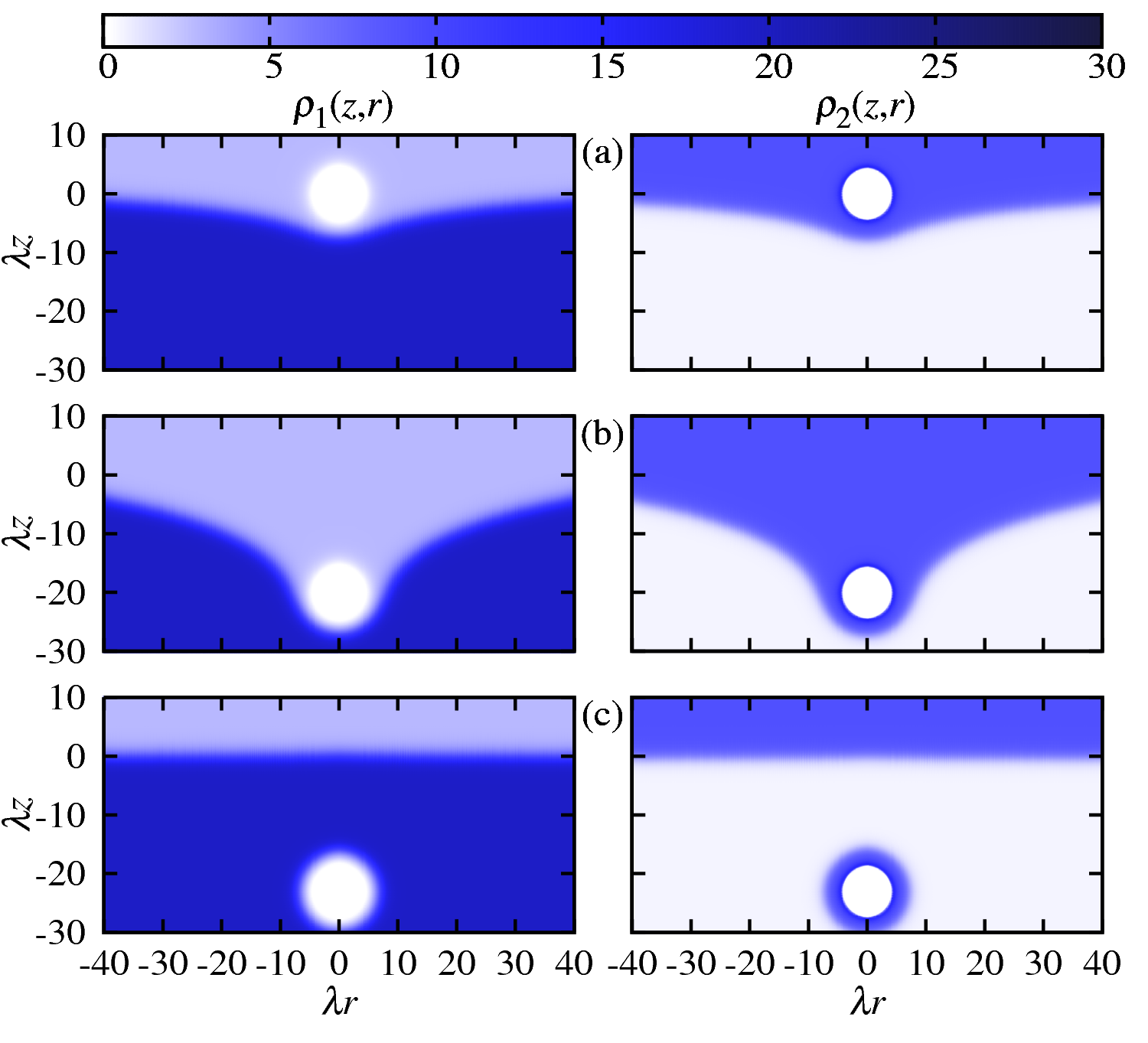}
 \caption{\label{fig:fi_profilesA}
(Colour Online) The solvent density profiles $\lambda^{-3}\rho_i(z,r)$ around colloid A at locations: (a) $z_0=0$, (b) $z_0=-20 \lambda^{-1}$ and (c) $z_0=-23 \lambda^{-1}$ in or near the fluid-fluid interface between co-existing phases X ($z<0$) and Y ($z>0$) in Fig.\ \ref{fig:phased}. The left hand density profiles are for solvent particles of species 1 and the right hand profiles for species 2. The grand potential corresponding to these density profiles is denoted by $\Box$ in Fig.\ \ref{fig:W_big_s34}.}
 \end{figure}

 \begin{figure}[t]
 \centering
 \includegraphics[width=8.5cm]{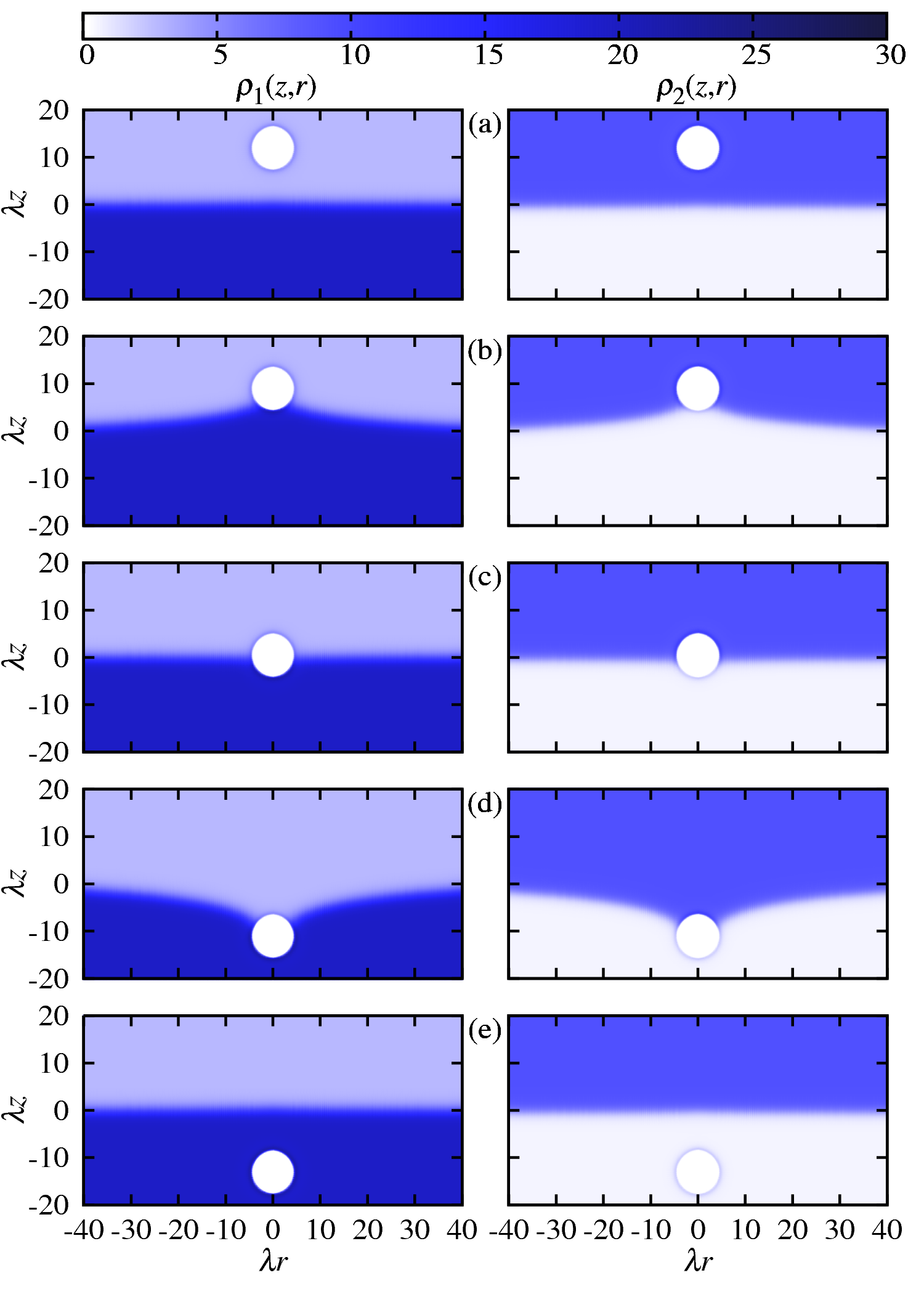}
 \caption{\label{fig:fi_profilesB}
(Colour online) Same as Fig.~\ref{fig:fi_profilesA} but now for colloid B at locations: (a) $z_0=12\lambda^{-1}$, (b) $z_0=9\lambda^{-1}$, (c) $z_0=0.5\lambda^{-1}$ (the global minimum of the grand potential energy and the equilibrium position for the colloid), (d) $z_0=-11\lambda^{-1}$ and (e) $z_0=-13\lambda^{-1}$. The grand potential corresponding to these density profiles is denoted by $\triangle$ in Fig.\ \ref{fig:W_big_s34}.}
 \end{figure}

\begin{figure}[t]
\centering
\includegraphics[width=8.5cm]{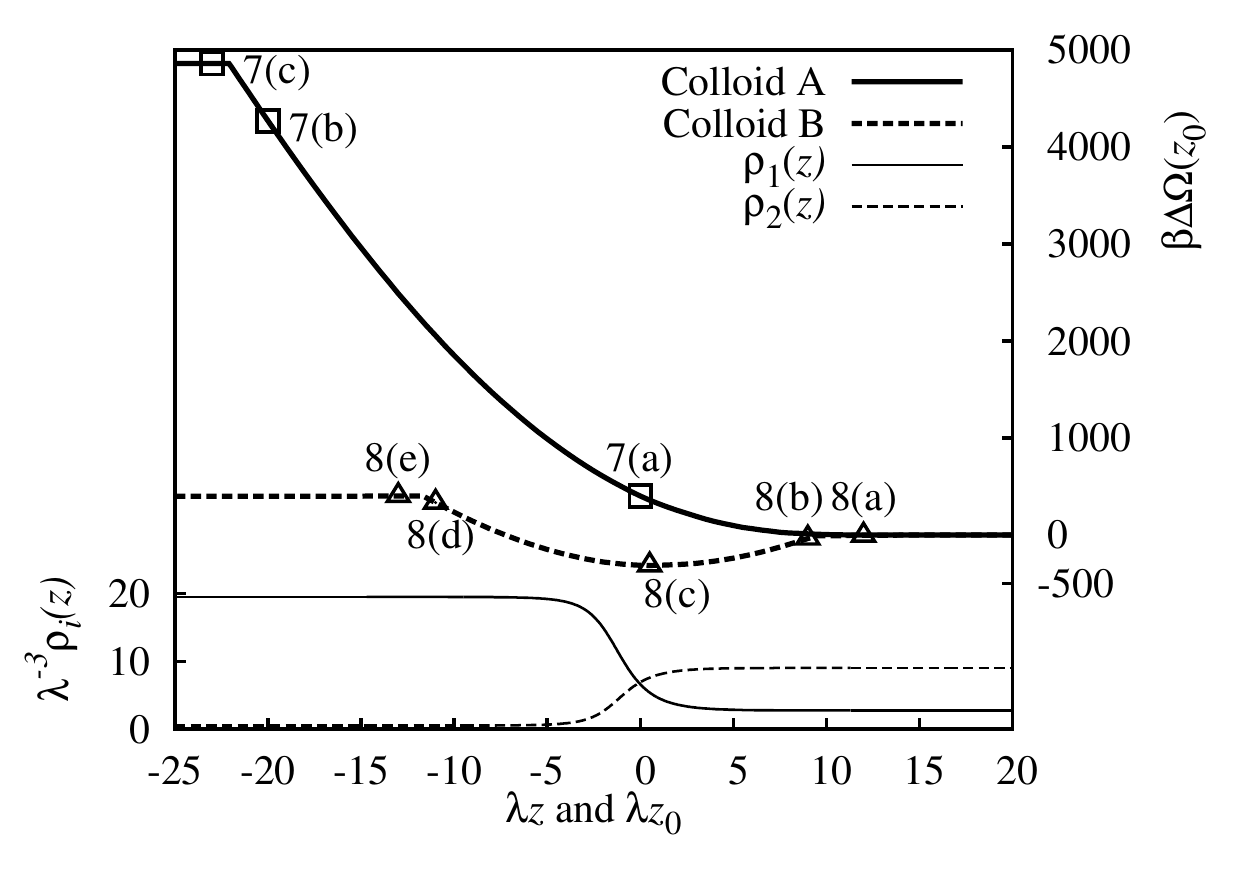}
\caption{\label{fig:W_big_s34}
The grand potential difference $\Delta\Omega(z_0)\equiv\Omega(z_0)-\Omega(\infty)$ for a single type A or type B colloid in the vicinity of the fluid-fluid interface between co-existing phases X ($z<0$) and Y ($z>0$) in Fig.\ \ref{fig:phased}; $z_0$ is the location of the center of the colloid. We also display the solvent density profiles $\rho_1(z)$ and $\rho_2(z)$ for the unperturbed interface (Fig.~\ref{fig:wall_profiles}(a)). The symbols $\Box$ and $\triangle$ carry labels referring to Figs.\ \ref{fig:fi_profilesA} or \ref{fig:fi_profilesB} which display the solvent density profiles around the colloid for that particular value of $z_0$. We see that colloid A strongly prefers to be in phase Y (rich in species 2) and so the minimum of $\Delta\Omega(z_0)$ for colloid A is at $z_0 \to \infty$. The `neutral' colloid B does not strongly prefer one phase over the other and there is a minimum in $\Delta\Omega(z_0)$ at $z_0=0.5\lambda^{-1}$, where the colloid sits within the interface; see Fig.\ \ref{fig:fi_profilesB}(c).}
\end{figure}

In Fig.\ \ref{fig:fi_profilesA} we display the density profiles for a single type A colloid at three different locations $z_0$; these lie on either side and within the fluid-fluid interface between coexisting phases X and Y. The plane of the fluid-fluid interface is set to be at $z=0$ with phase X being at $z<0$ and phase Y at $z>0$ as in Fig.~\ref{fig:wall_profiles}(a). When the colloid is far from the interface at $z_0 \to \pm \infty$ the colloid does not, of course, interact with the interface. Recall that colloid A strongly prefers to be in solvent phase Y (see Sec.\ \ref{subsec:single_colloid} above) and that when it is in phase X it is covered by a thick wetting film composed of the Y phase. This means that the global minimum of the grand potential for this system corresponds to $z_0 \to \infty$, when the colloid is deep in the bulk phase Y.  When the location of colloid A is shifted into the fluid-fluid interface the grand potential is minimized by maintaining a film rich in species 2 particles around the colloid. This leads to bending of the fluid-fluid interface to accommodate the colloid -- see Figs.~\ref{fig:fi_profilesA}(a) and \ref{fig:fi_profilesA}(b). As $z_0$ is further decreased, a point is reached where the grand potential cost from creating additional fluid-fluid interfacial area becomes equal to the grand potential for inserting the colloid into the bulk X phase. Beyond this point, the grand potential for the bent interface configuration, such as that in Fig.\ \ref{fig:fi_profilesA}(b), is greater than that of having the colloid covered by a wetting film that is disconnected from the fluid-fluid interface; see Fig.~\ref{fig:fi_profilesA}(c). Decreasing $z_0$ further does not change the grand potential. The SM potential (grand potential difference) $\Delta \Omega(z_0)\equiv\Omega(z_0)-\Omega(z_0 \to \infty)$ is displayed in Fig.\ \ref{fig:W_big_s34}. We see that $\Delta \Omega(z_0)$ increases continuously as $z_0$ is decreased until it reaches the point $z_0=-22.1 \lambda^{-1}$, where the film around the colloid detaches from the main fluid-fluid interface. This results in a discontinuity in the first derivative of $\Delta \Omega(z_0)$ at this point.

In Fig.\ \ref{fig:fi_profilesB} we display the density profiles for a single type B `neutral' colloid at several different locations $z_0$, i.e. either side and within the fluid-fluid interface between coexisting phases X and Y. The argument presented in Sec.\ \ref{sec:nondeform_inter} applies in this situation. Due to the fact that the colloid does not have a strong preference for either solvent phase the global minimum of the free energy occurs when the fluid-fluid interface intersects the colloid; the density profiles for this situation are displayed in Fig.\ \ref{fig:fi_profilesB}(c). If the colloid is moved to points on either side of the minimum at $z_0=0.5\lambda^{-1}$, the fluid-fluid interface remains connected to the colloid and has an increased area. This increase in interfacial area results in an increase in the grand potential. The solvent density profiles for such configurations are displayed in Fig.\ \ref{fig:fi_profilesB}(b) and \ref{fig:fi_profilesB}(d). Finally, as the colloid is moved even further from the interface a point is reached where the interface disconnects from the colloid. This is the case for the solvent density profiles displayed in Figs.\ \ref{fig:fi_profilesB}(a) and \ref{fig:fi_profilesB}(e). In Fig.\ \ref{fig:W_big_s34} we display $\Delta \Omega(z_0)$ obtained from these density profiles. At the two values of $z_0$ where the interface disconnects from the colloid, $z_0=-11.7\lambda^{-1}$ and $z_0=9.5\lambda^{-1}$, we find a discontinuity in the first derivative of $\Delta \Omega(z_0)$. Note that since we define $\Delta \Omega(z_0)\equiv\Omega(z_0)-\Omega(z_0 \to \infty)$, i.e.\ the grand potential when the colloid is at $z_0$ minus that for placing the colloid deep into the bulk of the Y phase, we find that $\Delta \Omega$ is zero for the density profiles in Fig.\ \ref{fig:fi_profilesB}(a). For the density profiles in Fig.\ \ref{fig:fi_profilesB}(e), i.e.\ inserting the colloid into the coexisting X phase, $\Delta \Omega=403k_BT$. That this value is greater than zero reflects the fact that the B colloids have slight preference for solvent phase Y over phase X. For the density profiles in Fig.\ \ref{fig:fi_profilesB}(c), at the minimum of the grand potential, $\Delta \Omega=-312k_BT$. We may compare this result with the value one would obtain from the macroscopic thermodynamic approach discussed in Sec.\ \ref{sec:nondeform_inter}, where we argued that in this situation $\Delta \Omega$ is given by Eq.\ \eqref{eq:omega_interface}, i.e.\ roughly equal to the change in the area of the interface between the two coexisting solvent phases when the colloid sits at the interface, $-\pi R^2$, multiplied by the fluid-fluid surface tension $\gamma$. Using the value for $\gamma$ given in Table \ref{tab:state_points} we find $\Delta \Omega=-\pi R^2 \gamma=-270k_BT$. One could also argue that because of the large values chosen for the parameters $b_i$ in Eq.\ \eqref{eq:testpart}, the effective radius $R^*$ of the colloid is somewhat larger than $R$. From inspecting the density profiles in Fig.\ \ref{fig:colloid_profiles}, we estimated $R^* \simeq 4.5 \lambda^{-1}$ (recall $R= 4 \lambda^{-1}$), giving an estimate for $\Delta \Omega\simeq-\pi R^{*2} \gamma=-350k_BT$, which is in reasonable agreement with the value we obtain from the microscopic DFT theory. What this shows is that one may use the crude approximation in Eq.\ \eqref{eq:omega_interface} to estimate roughly $\Delta \Omega$ for `neutral' colloids, even when the colloids are only one order of magnitude larger in size than the solvent particles.

Colloid C strongly prefers solvent phase X so when we insert colloid C into the fluid-fluid interface it behaves in the opposite way to colloid A, i.e. in the phase rich in species 2 (state-point Y) colloid C is covered by a thick layer rich in species 1. This means that the shape of the curve $\Delta \Omega(z_0)$ for colloid C (not displayed) is similar to the curve for colloid A in Fig.\ \ref{fig:W_big_s34}, but with the replacement $z_0\to -z_0$.

\subsection{Interaction between a single colloid and a wetting film adsorbed at a planar wall}
\label{subsec:hard_wall}

 \begin{figure}[t]
 \centering
\includegraphics[width=8.5cm]{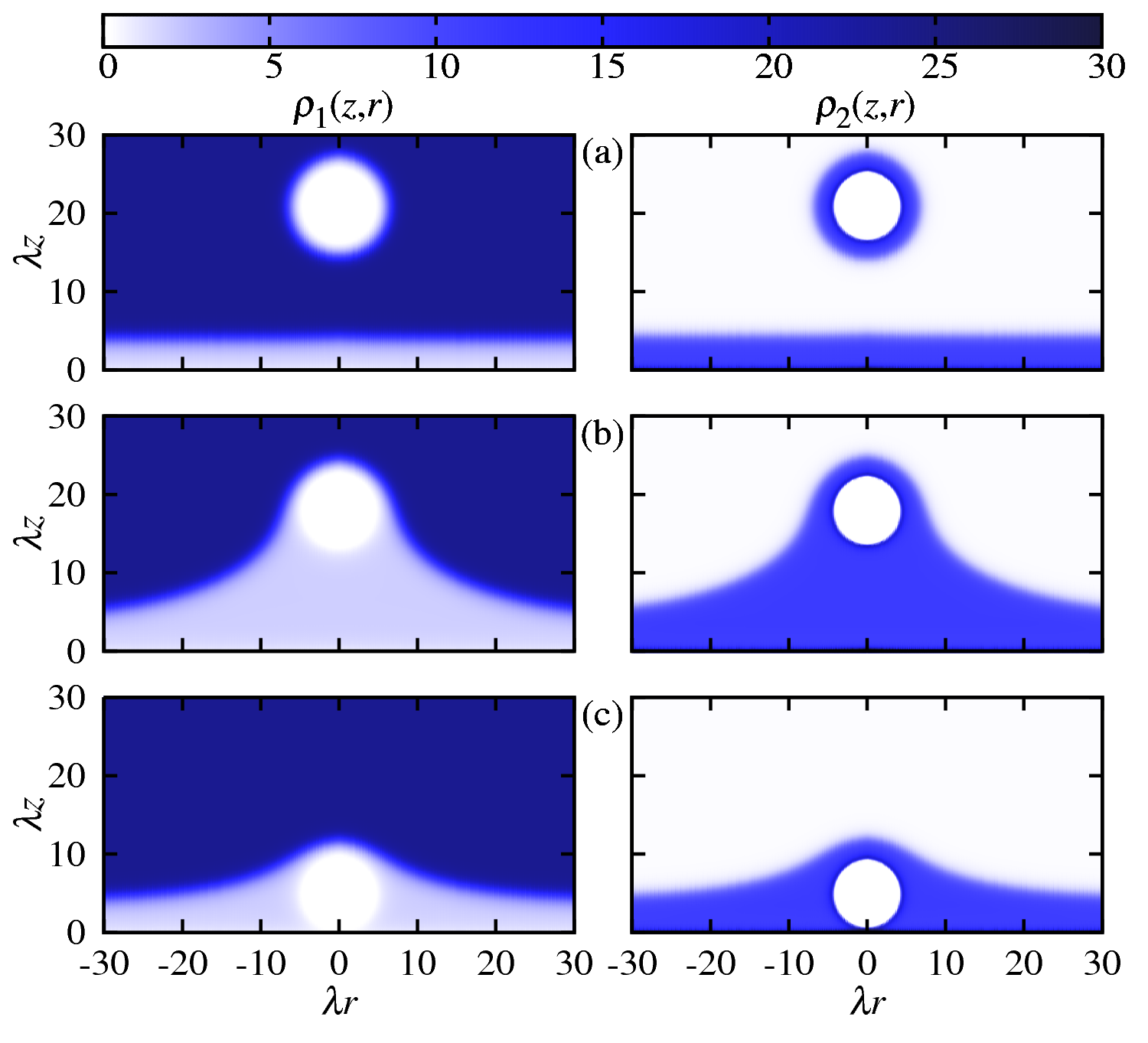}
\caption{\label{fig:dens_profs_we_A}
(Colour online) The solvent density profiles $\lambda^{-3}\rho_i(z,r)$ around a type A colloid located near a planar wall at $z=0$, with potentials given by Eq.\ \eqref{eq:yukwall}. In (a) the centre of the colloid is at $z_0=21\lambda^{-1}$, and the colloid is sufficiently far from the wall that the adsorbed films do not interact. In (b) $z_0=18 \lambda^{-1}$ and the wetting films covering the wall and the colloid connect. In (c) $z_0=5 \lambda^{-1}$, and the colloid is close to contact with the wall. The bulk solvent is at state point W in Fig.\ \ref{fig:phased} and is rich in species 1. The left hand profiles are for species 1 and the right hand profiles for species 2. The grand potential for these density profiles is denoted by $\Box$ in Fig.\ \ref{fig:part_wall_omega_bar}.}
  \end{figure}

 \begin{figure}[t]
 \centering
\includegraphics[width=8.5cm]{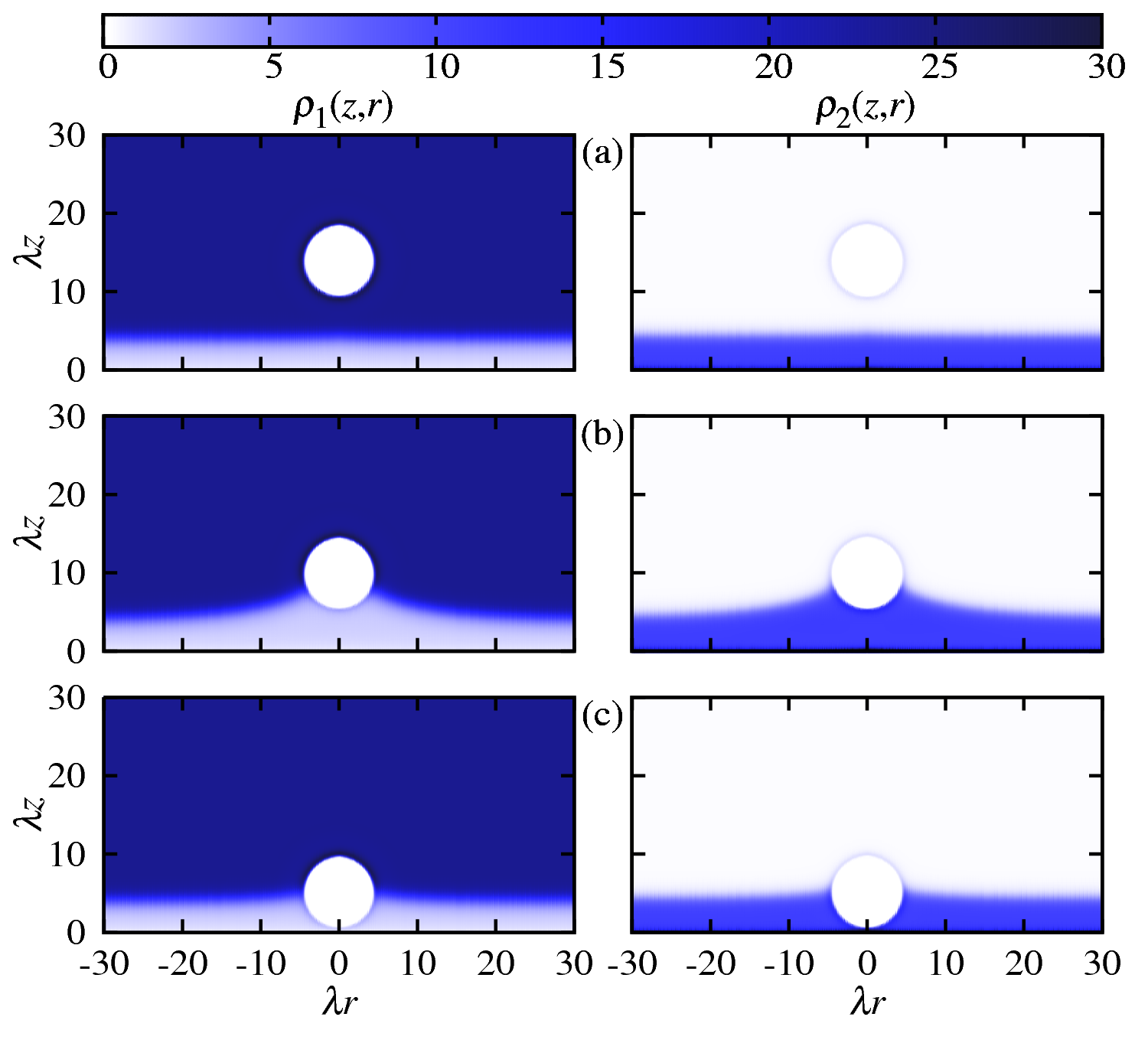}
\caption{\label{fig:dens_profs_we_B}
(Colour online) Same as Fig.~\ref{fig:dens_profs_we_A} but for a type B colloid. In (a) $ z_0=14\lambda^{-1}$ and the colloid is sufficiently far from the wetting film at the wall that these do not interact. In (b) $z_0=10\lambda^{-1}$ and the film at the wall bends so that the interface intersects the colloid. In (c) $z_0=5.2\lambda^{-1}$ and the colloid is nearly in contact with the wall. The grand potential corresponding to these density profiles is denoted by $\triangle$ in Fig.\ \ref{fig:part_wall_omega_bar}.}
  \end{figure}

 \begin{figure}[t]
 \centering
\includegraphics[width=8.5cm]{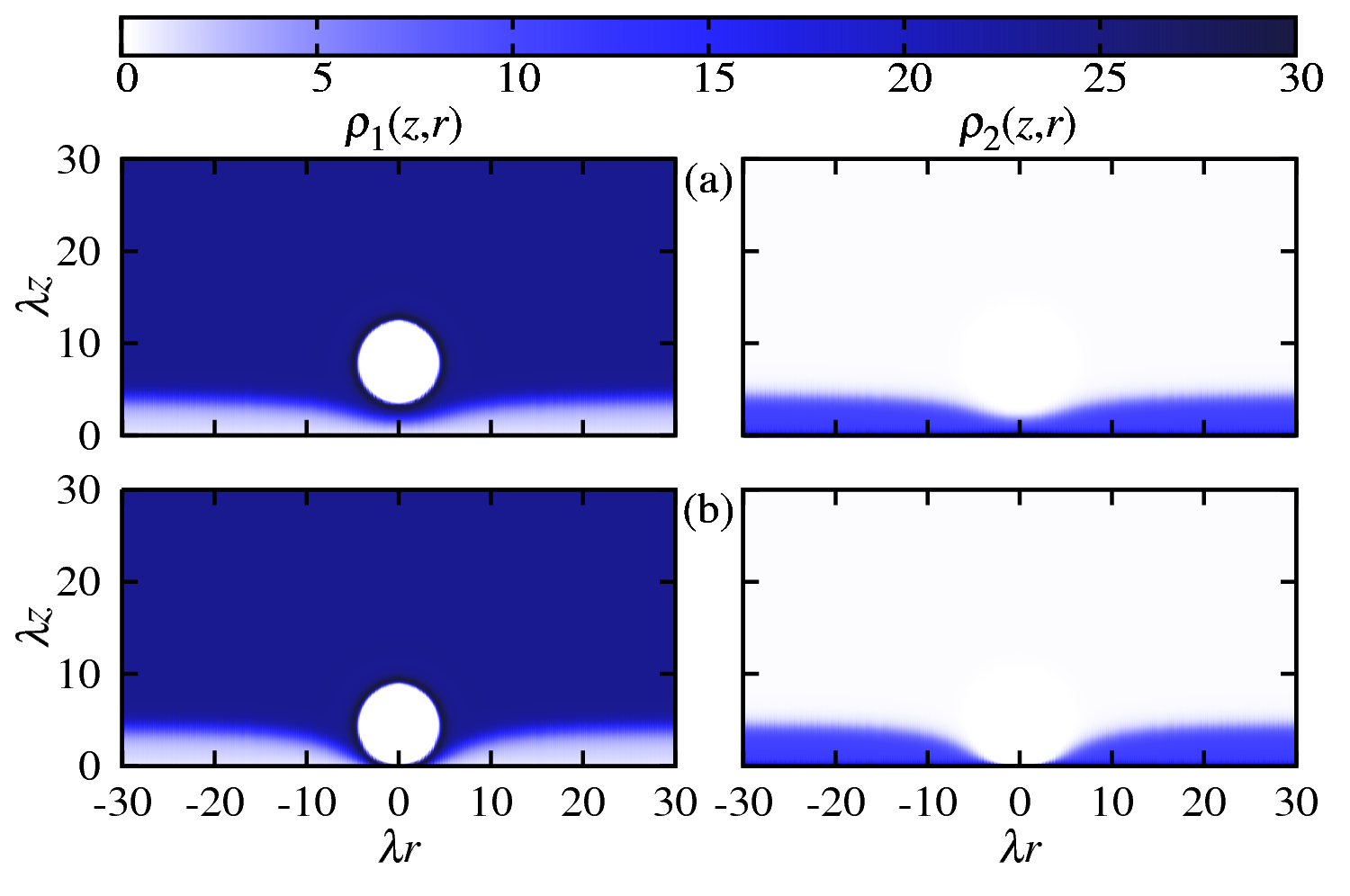}
\caption{\label{fig:dens_profs_we_C}
(Colour online) Same as Fig.~\ref{fig:dens_profs_we_A} but for a type C colloid. In (a) $z_0=8\lambda^{-1}$. Since the colloid strongly prefers to be in the bulk solvent phase the wetting film at the wall thins so that the colloid can maintain a layer of the bulk phase around itself as it approaches the wall. In (b) $z_0=4.5\lambda^{-1}$ and the colloid is almost in contact with the wall. The grand potential corresponding to these density profiles is denoted by $\diamond$ in Fig.\ \ref{fig:part_wall_omega_bar}.}
  \end{figure}

\begin{figure}[t]
\centering
\includegraphics[width=8.5cm]{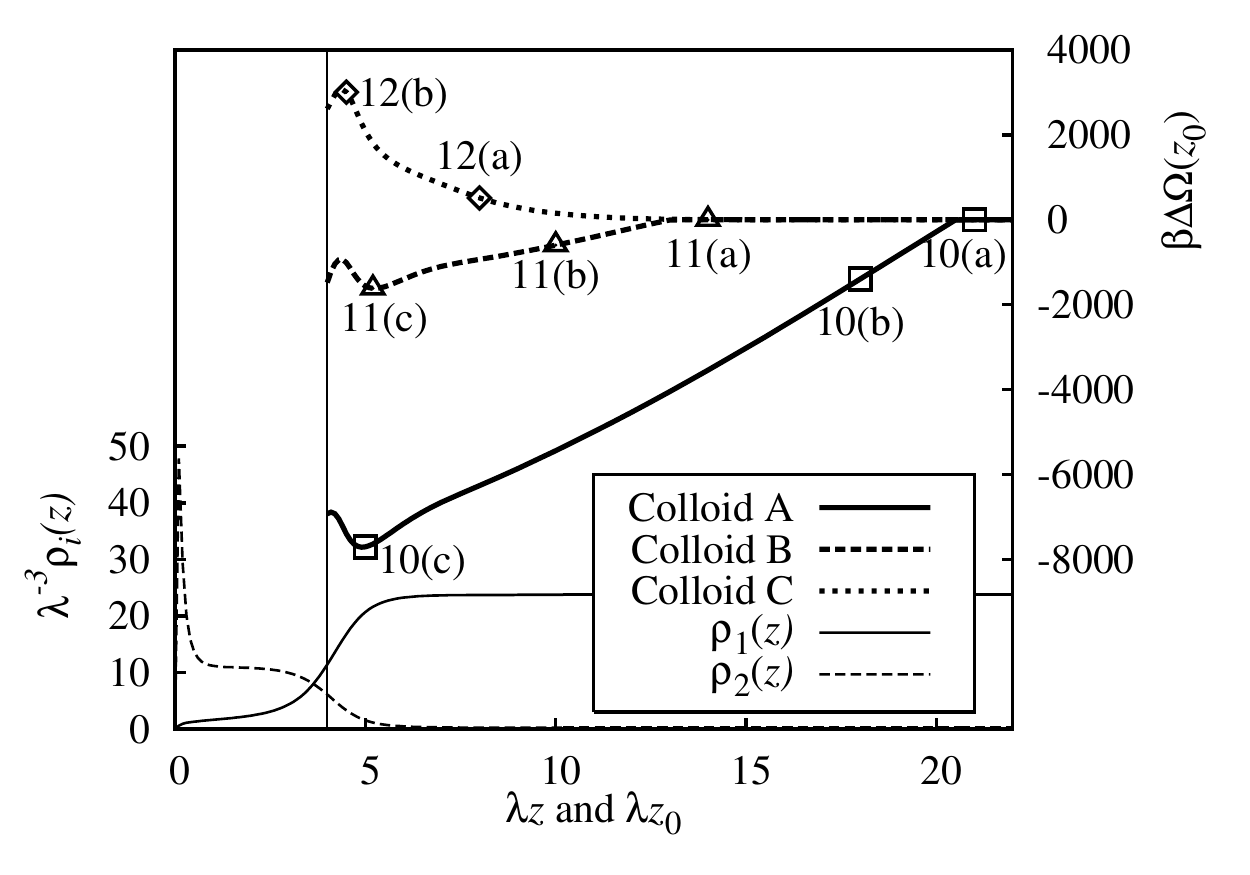}
\caption{\label{fig:part_wall_omega_bar}
The grand potential difference $\Delta\Omega(z_0)\equiv\Omega(z_0)-\Omega(\infty)$ (solvent mediated potential) for a single type A, B, or C colloid. $z_0$ is the distance of the centre of the colloid from a planar wall at $z=0$ with potentials given by Eq.\ \eqref{eq:yukwall}. We also display the solvent density profiles $\rho_i(z)$ near the wall when no colloid is present (Fig.~\ref{fig:wall_profiles}(b)); the wall is covered by a thick wetting film rich in species 2. Symbols $\Box$, $\triangle$ and $\diamond$ carry labels referring to Figs.\ \ref{fig:dens_profs_we_A} -- \ref{fig:dens_profs_we_C} which display the solvent density profiles around the colloid for these particular values of $z_0$. The bulk solvent is at state point W in Fig.~\ref{fig:phased}, and the vertical line is at $z_0=4\lambda^{-1}$, the colloid radius.}
\end{figure}

In Fig.\ \ref{fig:dens_profs_we_A} we display the solvent density profiles around a single type A colloid positioned at three different distances $z_0$ from a planar wall, located at $z=0$, with potentials given by Eq.\ \eqref{eq:yukwall}. As the bulk solvent is at state point W in Fig.\ \ref{fig:phased}, which is near to phase coexistence, the wall is covered by a thick wetting film of the coexisting phase rich in species 2. In the bulk fluid, away from the wall, colloid A is also covered by a thick wetting film. When the colloid is far from the wall, the films do not interact -- see for example the density profiles in Fig.\ \ref{fig:dens_profs_we_A}(a). As the colloid is brought closer to the wall the solvent density profiles change abruptly from a configuration such as that displayed in Fig.\ \ref{fig:dens_profs_we_A}(a) to a configuration where there is a bridge of the wetting phase connecting the colloid to the wall; see Fig.\ \ref{fig:dens_profs_we_A}(b). The position of the colloid when the bridge forms is $z_0=z_t=19.8\lambda^{-1}$. As $z_0$ is decreased further, the colloid remains within the wetting film all the way to contact with the wall at $z_0=4\lambda^{-1}$. In Fig.\ \ref{fig:part_wall_omega_bar} we display the solvent mediated potential (grand potential difference) $\Delta \Omega(z_0)\equiv \Omega(z_0)-\Omega(z_0\to \infty)$ calculated from these density profiles. We see that for $z_0<z_t$, $\Delta \Omega(z_0)$ is negative indicating that the effective interaction between colloid A and the wall is strongly attractive. When the colloid is close to contact with the wall $\Delta \Omega(z_0)\simeq-7050k_BT$. The primary origin of this attraction lies in the fact that when $z_0<z_t$ the area of the interface between the bulk fluid and the wetting film covering the wall and the colloid is less than when the colloid is in the bulk away from the wall at $z_0>z_t$. The abrupt change in the density profiles at $z_0=z_t$ manifests as a discontinuity in the first derivative of $\Delta \Omega(z_0)$, i.e. a jump in the SM force between the colloid and the wall at $z_0=z_t$.

In Fig.\ \ref{fig:dens_profs_we_B} we display the solvent density profiles around a single type B colloid positioned at three different distances $z_0$ from the wall. Colloid B is not covered by a wetting film in the bulk and as it is brought closer to the wall there is an abrupt change in the solvent density profiles at $z_0 =z_t= 13.3\lambda^{-1}$, from a configuration where the wetting film covering the wall is almost completely unperturbed,to a configuration where the wetting film extends to meet the colloid, so that the interface between the bulk fluid phase and the wetting film is anchored to the colloid. As $z_0$ is decreased further this interface remains connected to the colloid; see Fig.\ \ref{fig:dens_profs_we_B}(b) and (c). In Fig.\ \ref{fig:part_wall_omega_bar} we display the grand potential difference $\Delta \Omega(z_0)\equiv \Omega(z_0)-\Omega(z_0\to \infty)$ calculated from these density profiles. We see that at $z_0=z_t$ there is a discontinuity in the first derivative of $\Delta\Omega(z_0)$, and that $\Delta \Omega(z_0)$ is negative for $z_0<z_t$, indicating that the effective interaction between colloid B and the wall is attractive. Close to contact $\Delta \Omega(z_0)\simeq-1500 k_BT$, so the effective interaction between colloid B and the wall is much less strongly attractive than that between colloid A and the same wall.

In Fig.\ \ref{fig:dens_profs_we_C} we display the solvent density profiles around a single type C colloid positioned at two different distances $z_0$ from the wall. Colloid C strongly prefers to be in the bulk solvent phase, which is poor species 2, rather than in the phase wetting the wall, which is rich in species 1. As colloid C is brought close to the wall it moves into the vicinity of the wetting film covering the wall and because colloid C prefers the bulk phase, the wetting film at the wall is thinned (and eventually expelled) in the region between the colloid and the wall. The density profiles change continuously as $z_0$ is varied. In Fig.\ \ref{fig:part_wall_omega_bar} we display the grand potential difference $\Delta \Omega(z_0)$. As $z_0$ is decreased $\Delta\Omega(z_0)$ increases smoothly as the grand potential for having the bulk solvent phase close to the wall is greater than for having the wall covered by a uniform thick wetting film of the coexisting phase. The fact that near the wall $\Delta \Omega(z_0)$ is positive indicates that the effective interaction between colloid C and the wall is repulsive, in marked contrast to colloids A and B.

\section{Discussion and Concluding Remarks}
\label{sec:discuss}
We have investigated the behavior of model colloids inserted into a binary mixture of softcore solvent particles that exhibits fluid-fluid demixing. We considered solvent statepoints at, or close to, fluid-fluid coexistence where the solvent correlation functions are reasonably well described by a simple RPA DFT. By varying the colloid-solvent interaction parameters we were able to investigate the colloid's affinity for insertion into one of the bulk coexisting phases. For type A and C colloids we showed that when these are in isolation in the unfavored bulk solvent phase they adsorb a thick `wetting' film of the favored coexisting phase. Type B colloid has no strong preference for either bulk phase.

Using a `brute-force' DFT approach that models the colloids by an external potential acting on the solvent particles we calculated the density profiles of the solvent for the following scenarios: (i) a pair of colloids of the same type immersed in the bulk, (ii) a colloid in a planar fluid-fluid interface, and (iii) a colloid near a planar wall that adsorbs a thick wetting film. By calculating the grand potential of the solvent we were able to determine the SM interaction in each case. We found that the presence (or absence) of a thick adsorbed film around the colloids in isolation determines the behaviour of the system in the different scenarios.

When two colloids, both with a thick adsorbed film, are brought together then the films can connect together abruptly to form a fluid bridge. A similar phenomenon occurs when a colloid is brought close to a wall. The formation of a fluid bridge gives rise to a strong, long-ranged SM attraction. If there are no thick wetting films then the SM potential is much less attractive and much shorter ranged. Furthermore, when the wall and colloid have very different solvent affinities then the SM interaction can be repulsive.

For a type B colloid inserted in the fluid-fluid interface the effective colloid-interface interaction has an attractive well implying that such `neutral' colloids would become trapped in the interface. On the other hand, colloids A or C that strongly prefer to be in one of the bulk phases distort the interface, and ultimately experience a repulsive interaction with the interface.

In our mean-field DFT treatment we find an abrupt, first-order like transition for many of the scenarios studied, i.e. the SM force is discontinuous when a fluid bridge forms.  However, since a finite number of particles are involved in forming the bridging film and in the onset of interface distortion, these cannot be true phase transitions. In reality fluctuation (finite size) effects must result in a rounding of the discontinuity in the SM force.~\cite{bauer2000wie} The authors of Ref.~\cite{archer2005smi} suggested that a crude estimate for the width of this rounding, $\delta z_{0,\mathrm{br}}$, can be obtained by considering that fluctuations should only be relevant when $|\Delta \Omega^{\mathrm{Bridged}}(z_0)-\Delta \Omega^{\mathrm{Unbridged}}(z_0)|\lesssim k_BT$ where $\Delta \Omega^{\mathrm{(Un)Bridged}}(z_0)$ is the grand potential on the (un)bridged side of the transition. Using this crude criterion we find that the transition is smeared over a length $\delta z_{0,\mathrm{br}}/z_{0,\mathrm{br}}\sim10^{-3}$. This is similar to the value estimated by Archer \etal~\cite{archer2005smi} in their DFT studies of bridging between two big solute particles in a Gaussian core model of a binary solvent.

We showed that for two colloids in bulk the strength and range of the SM interaction is reasonably well described by a simple (capillarity) model for the shape of the liquid bridge formed between the two colloids, namely Eqs.~\eqref{eq:Omega_ex_beta_wet_2} and \eqref{eq:S_approx}. The only inputs to this theory are the surface tension of the planar interface between the two solvent phases and the thickness of the adsorbed `wetting' film. These quantities can be obtained from simpler DFT calculations since they depend only on profiles that vary in one dimension. For the neutral colloid immersed in the planar fluid-fluid interface we calculated the depth of the attractive well in the effective colloid-interface interaction and found this to be reasonably close to the value predicted from a very simple model that considered only the surface of the colloid in an unbending interface, Eq.~\eqref{eq:omega_interface}. It is interesting that these macroscopic (capillarity) approximations appear to have some validity in the case where the colloids are only one order of magnitude larger than the solvent particles. Furthermore, more systematic comparisons of the results of full DFT calculations with those based on macroscopic approaches would be valuable in ascertaining the limitations of the latter for nanoparticles.
 
We have deliberately limited the scope of this study and have not conducted an extensive investigation of the entire phase diagram, nor of the full set of parameters characterizing the solvent-colloid and solvent-wall interactions. It is likely that the presence of a pre-wetting transition would lead to additional features in the SM potentials. Furthermore, we have only investigated problems where the density profiles exhibit cylindrical symmetry. The interactions between three hard-sphere colloids in a bulk hard-sphere solvent have been investigated both with brute force DFT~\cite{melchionna2000tdf}, and the particle insertion method~\cite{goulding2001act}, and an obvious extension of the present work would be to consider the interactions between three colloids in the bulk, or two colloids in the presence of a fluid interface, or near a planar wall.  This would necessarily require a full three dimensional DFT computation, which in turns brings its own complications. The density profiles can no longer be easily calculated with sufficient precision on a desktop computer but instead one might use a number of computers in parallel.~\cite{frink2000tat} Furthermore, particular care must be taken in establishing that the density profiles are the true equilibrium profiles and that any abrupt jumps correspond to the equilibrium phase transitions, i.e. the grand potential must be calculated very precisely; this poses considerable numerical challenges.

Our approach can be compared to investigations using both molecular dynamics and Monte Carlo simulations. Many of the studies mentioned in the Introduction used Lennard-Jones or similar potentials for the solvent-solvent and solvent-colloid interactions, and further work based on the DFT approach should attempt to treat such models. On the other hand, a particular advantage of the present simple model is that the ratio between the colloid and solvent particle sizes is sufficiently large that we are able to obtain a thick film adsorbed around modest sized (soft core) colloids. Existing simulation studies using Lennard-Jones potentials have used a maximum colloid diameter of 10 times the solvent particle diameter.~\cite{bresme1999csw} This is not sufficient to adsorb a thick wetting film for hard core colloids.

So far in this paper we have not indicated the form of the bare interaction between the large colloids, nor the interaction between the colloids and the wall, since these have no bearing on the SM potentials. We suggest that the bare colloid-colloid potential has a hard-core of diameter $8\lambda^{-1}$ and that the potential decreases rapidly outside this hard-core. Then the effect of the bridging film is to still induce a strong, long-ranged attraction between the colloids for suitable solvent statepoints. Similarly, the bare wall-colloid potential should diverge rapidly at $z=4\lambda^{-1}$, and for suitable solvent state-points the long-ranged attraction should be retained. Finally, although our model does not correspond directly to an experimental situation it has been suggested on general grounds that light scattering experiments could be used to investigate the value of the second virial coefficient, $B_2$ of colloids in bulk solvents.~\cite{archer2005smi} Rapid changes of $B_2$ upon changing the solvent state-point might signal the aggregation of colloids driven by the formation of fluid bridges.

\begin{acknowledgments}
We are grateful to George Jackson, Jeroen Van Duijneveldt and Matthias Schmidt for illuminating discussions and to Matthias Schmidt for the use of computing facilities. PH thanks EPSRC and AJA thanks RCUK for financial support.
\end{acknowledgments}


\begin{thebibliography}{10}
\bibitem{footnote1}
Note that in Eq.\ \eqref{eq:Omega_particle} we have implicitly assumed that the colloid is hard (i.e.\ the potentials $\phi_i^{sc}(r)$ that it exerts on the solvent particles have the property $\phi_i^{sc}(r \leq R)=\infty)$ and therefore choosing the Gibbs dividing surface at $r=R$ is the natural choice. One can, of course, define the dividing surface at any value of $r$ without changing $\Omega$. The choice of location for the dividing surface does, however, affect the values for the available volume and $\gamma_{p \alpha}$ used in Eq.\ \eqref{eq:Omega_particle}.
\end{thebibliography}
\end{document}